\documentclass[aps,pre,twocolumn,showpacs,superscriptaddress]{revtex4}

\usepackage{amsfonts,amssymb,amsmath,times}
\usepackage{graphicx,color,enumerate}

\begin{document}

\title{Geometric and dynamic perspectives on phase-coherent and noncoherent chaos}

\author{Yong Zou}
    \affiliation{Potsdam Institute for Climate Impact Research, P.~O. Box 601203, 14412 Potsdam, Germany}
	\affiliation{Department of Electronic and Information Engineering, Hong Kong Polytechnic University, Hung Hom, Kowloon, Hong Kong}
    \affiliation{Department of Physics, East China Normal University, 200062 Shanghai, China}
\author{Reik V. Donner}
    \affiliation{Potsdam Institute for Climate Impact Research, P.~O. Box 601203, 14412 Potsdam, Germany}
\author{J\"urgen Kurths}
    \affiliation{Potsdam Institute for Climate Impact Research, P.~O. Box 601203, 14412 Potsdam, Germany}
    \affiliation{Department of Physics, Humboldt University Berlin, Newtonstr.~15, 12489 Berlin, Germany}
 	\affiliation{Institute for Complex Systems and Mathematical Biology,
 	University of Aberdeen, Aberdeen AB243UE, United Kingdom}

\date{December 22, 2011}

\begin{abstract}

Statistically distinguishing between phase-coherent and noncoherent chaotic dynamics from time series is a contemporary problem in nonlinear sciences. In this work, we propose different measures based on recurrence properties of recorded trajectories, which characterize the underlying systems from both geometric and dynamic viewpoints. The potentials of the individual measures for discriminating phase-coherent and noncoherent chaotic oscillations are discussed. A detailed numerical analysis is performed for the chaotic R\"ossler system, which displays both types of chaos as one control parameter is varied, and the Mackey-Glass system as an example of a time-delay system with noncoherent chaos. Our results demonstrate that especially geometric measures from recurrence network analysis are well suited for tracing transitions between spiral- and screw-type chaos, a common route from phase-coherent to noncoherent chaos also found in other nonlinear oscillators. A detailed explanation of the observed behavior in terms of attractor geometry is given.

\end{abstract}

\maketitle

\textbf{
Oscillatory processes can be frequently observed in natural and technological systems. Often, the corresponding dynamics is not strictly periodic, but shows more complex temporal variability patterns characterized by a fast divergence of trajectories with arbitrarily close initial conditions~\cite{MTS1,BifurTheory,Lichtenberg_Lieberman_regular}. There are numerous examples of such chaotic oscillators for which long-term predictions of amplitudes and phases are not possible. Therefore, studying their phase dynamics has recently attracted particular interest, e.g., regarding the process of phase synchronization between different coupled systems~\cite{Rosenblum1996,Pikovsky_Kurths_synchr}. However, most existing methods suitable for this purpose require the explicit definition of an appropriate phase variable, which can become a non-trivial problem in the case of noncoherent chaotic oscillations. Therefore, studying the phase coherence properties of chaotic systems has become an important problem in both theoretical and experimental studies~\cite{wickramasinghe_chaos_2010}. In this work, we propose some methods based on the concept of recurrences in phase space, which allow studying complementary aspects of chaotic oscillators relating to the geometric structure of, and the dynamics on the attractor. Specifically, we derive a detailed characterization of changes of the geometric structure of complex systems in phase space with varying control parameter, which accompany transitions from phase-coherent to noncoherent dynamics.
}

\section{Introduction}

In the last decades, the complexity of chaotic oscillators has been widely characterized by a variety of different quantities inspired from nonlinear dynamical systems theory~\cite{Kantz97,Sprott2003}. Lyapunov exponents~\cite{Wolf,LyapuEx} describe the characteristic time-scale associated with the finite-time exponential divergence of nearby chaotic orbits and, thus, relate directly to the predictability horizon of the dynamics. Fractal dimensions and entropies measure the structural complexity of the underlying attractor, often based on concepts from information theory. 

\begin{figure}
  \centering
  \includegraphics[width=\columnwidth]{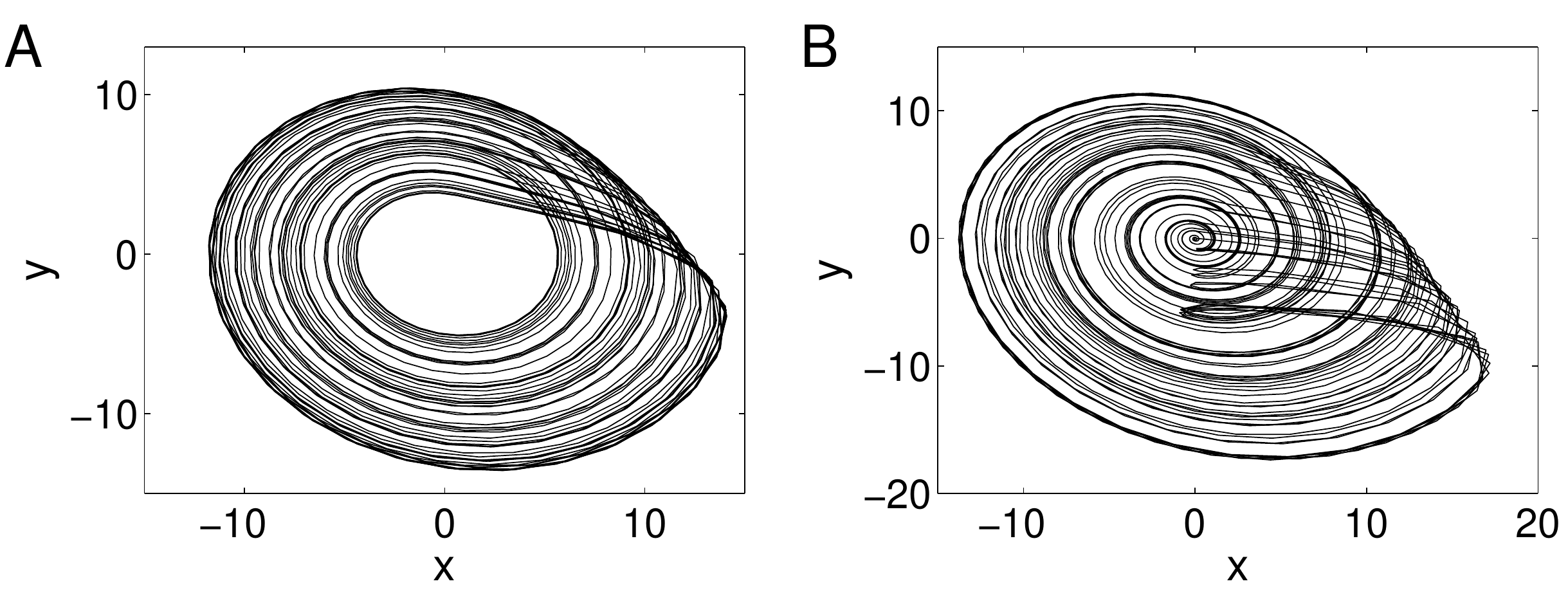}
  \caption{\small {Two-dimensional projection of a part of the trajectory of the R\"ossler system (Eq.~(\ref{eq_roessler})) in the (A) PC ($a=0.165$) and (B) NPC (funnel) regime ($a=0.265$).}
    \label{fig:ros_phasespace} }
\end{figure}

In contrast to the aforementioned concepts, in many situations, one is interested in explicitly characterizing the phase dynamics of the recorded nonlinear oscillations. However, depending on the structural properties of the chaotic oscillations under study, it may be difficult to assign a well-defined phase variable to the observed dynamics. This problem predominantly occurs in the presence of noisy oscillations, however, also in the fully deterministic case, one frequently observes oscillations without a distinct center of rotation in phase space, e.g., in the funnel regime (see Fig.~\ref{fig:ros_phasespace}B) of the R\"ossler system~\cite{Roessler1976}
\begin{equation}
\begin{split}
\dot{x} &= -y - z, \\
\dot{y} &= x + a y, \\
\dot{z} &= 0.4 + z (x - 8.5).
\end{split}
\label{eq_roessler}
\end{equation}
\noindent
In case of such \textit{noncoherent} oscillations, the appropriate definition and analysis of the phase dynamics is challenging. Therefore, given the rising number of examples of real-world chaotic oscillators, the problem of automatically distinguishing between phase-coherent (PC) and noncoherent (NPC) chaos is of practical relevance. Traditionally, this problem has been considered by studying the phase diffusion properties of the system under study~\cite{Pikovsky_Kurths_synchr}. However, in order to apply this conceptual idea, an appropriate phase variable has to be defined in advance.

In this work, we propose an alternative approach based on the recurrence properties of a dynamical system's trajectory in phase space for quantitatively characterizing whether or not an observed chaotic dynamics is phase-coherent. In contrast to the explicit study of phase diffusion, the corresponding concepts do \textit{not} rely on an explicit definition of a phase variable. We emphasize that this fact has already motivated using recurrence-based properties for studying synchronization processes of coupled NPC  oscillators~\cite{Mamen_epl_2005,Marwan_report_2007} and time-delay systems~\cite{Senthilkumar2010,Suresh2010}.

Generally, recurrence properties can be conveniently analyzed by using \textit{recurrence plots} (RPs)~\cite{Marwan_report_2007}, originally introduced in the seminal work by Eckmann \textit{et~al.}~\cite{Eckmann1987}, which provide an intuitive visualization of the underlying temporal structures. For this purpose, one defines the \textit{recurrence matrix} $R_{i,j}$ as a binary representation of whether or not pairs of observed state vectors on the same trajectory are mutually close in phase space. Given two state vectors $\mathbf{x}_i$ and $\mathbf{x}_j$ (where $i$ and $j$ are time indices), this proximity is most commonly characterized by comparing the length of the difference vector between $\mathbf{x}_i$ and $\mathbf{x}_j$ with a prescribed maximum distance $\varepsilon$, i.e.,
\begin{equation}
R_{i,j}(\varepsilon)=\Theta(\varepsilon-\|\mathbf{x}_i-\mathbf{x}_j \|),
\label{eq_defrp}
\end{equation}
\noindent
where $\Theta(\cdot)$ is the Heaviside function and $\| \cdot\|$ a norm (e.g., Euclidean, Manhattan, or maximum norm). In this work, we will specifically chose the maximum norm for defining distances in phase space, since it has lower computational demands than other norms. However, the choice of a different norm would not change the presented results qualitatively. The properties of RPs have been intensively studied for different kinds of dynamics~\cite{Marwan_report_2007}, including periodic, quasiperiodic~\cite{Zou_quasiperiod,Zou_chaos_2007,Zou_2008}, chaotic, and stochastic dynamics~\cite{Rohde2008,Marwan2009Comment}. 

It has been shown that, among other features, the length distributions of diagonal and vertical structures in RPs can be used for defining a variety of measures of complexity, which characterize properties such as the degree of determinism or laminarity of the system~\cite{Zbilut1992,Webber,Trulla1996,Physio1}. The resulting toolbox of recurrence quantification analysis (RQA) has been widely applied for studying phenomena from various scientific disciplines~\cite{Marwan_report_2007,Marwan2008}. In this work, however, we will utilize some complementary conceptual approaches also based on RPs, which do not belong to the set of classical RQA measures. Based on Eq.~(\ref{eq_defrp}), we will discuss properties based on the recurrence time statistics and so-called $\varepsilon$-recurrence networks (RNs). The underlying methodological concepts are briefly described in Sec.~\ref{sec:methods} and subsequently applied to two realizations of the R\"ossler system in PC and NPC regime, respectively. Following the results obtained for this example, potential new statistical indicators for phase coherence based on the recurrence properties of the underlying system are introduced in Sec.~\ref{sec:pcmeasures} and compared to other established as well as novel measures based on phase diffusion and Poincar\'e return times, respectively. Application to a complete bifurcation sequence of the R\"ossler system in Sec.~\ref{sec:roessler} demonstrates the feasibility of the recurrence-based approaches. The geometric consequences of the transition from PC (spiral-type) to NPC (screw-type) chaos and their impact on the recurrence properties are discussed. As a second example, the behavior of the recurrence based measures is illustrated for the Mackey-Glass system~\cite{Mackey1977} in a parameter range including transitions between periodic and NPC chaotic behavior~\cite{Farmer1982}.

\section{Methods}\label{sec:methods}

\subsection{Recurrence time statistics}\label{sec_rts}

Complementary to RQA, another natural way for characterizing the recurrence
properties of dynamical systems in phase space is statistically evaluating the
distribution of recurrence times (RTs), which has been applied to both chaotic
and stochastic
systems~\cite{Balakrishnan1997,Gao1999,Balakrishnan2000,Altmann2004}. In
contrast to return times with respect to a fixed Poincar\'e surface, recurrence
times refer to the time intervals after which the trajectory enters the
$\varepsilon$-neighborhood of a previously visited point in phase space. Gao
\textit{et~al.}~\cite{Gao2003} demonstrated that similar to some line-based RQA
measures, characteristics based on the RT distributions $p(\tau)$ can be used
for detecting subtle dynamical transitions, which motivated using a
corresponding approach for testing against
stationarity~\cite{Rieke2004,Rieke2004b}. Besides their immediate importance for
studies on extreme events~\cite{Altmann2005}, recurrence times have also proven their potential for
the estimation of dynamical invariants such as the information
dimension~\cite{Gao1999} and the Kolmogorov-Sinai entropy~\cite{Baptista2010}.

Given a RP, RTs can be identified as the lengths of non-interrupted vertical (or
horizontal, since the recurrence matrix is symmetric) ``white lines'' that do
not contain any recurrence (i.e., no pair of mutually close state vectors). More
precisely, such a white line of length $\tau$ starts at the position $(i,j)$ in
the RP if~\cite{Thiel03}
\begin{equation} \label{wvl_eq}
R_{i, j+m} = \begin{cases}
 1 & \text {if} \; m = -1, \\
 0 & \text {for} \; m \in \{0, \ldots, \tau-1 \}, \\
 1 & \text {if} \; m = \tau.
\end{cases}
\end{equation}
In order to see this, for all times $k=j-1, \ldots, j+\tau$, the values
$\mathbf{x}_k$ on the trajectory are compared with $\mathbf{x}_i$. Then, the
structure given by Eq.~(\ref{wvl_eq}) can be interpreted as follows: At time
$k=j-1$, the trajectory falls into an $\varepsilon$-neighborhood of
$\mathbf{x}_i$. Then, for $k=j, \ldots, j+\tau-1$, it moves further away from
$\mathbf{x}_i$ than a distance $\varepsilon$, until at $k=j+\tau$, it returns to
the $\varepsilon$-neighborhood of $\mathbf{x}_i$ again. Hence, given a uniform
sampling of the trajectory in the time domain, the length of the line is
proportional to the time that the trajectory needs to return $\varepsilon$-close
to $\mathbf{x}_i$. Going beyond the concept of \textit{first-return times}, the
ensemble of all recurrences to the $\varepsilon$-neighborhood of $\mathbf{x}_i$
induces a RT distribution for this specific point. Combining this information
for all available points $\mathbf{x}_i$ in a given time series (i.e.,
considering the lengths of all white lines in the RP), one obtains the RT
distribution $p(\tau)$ associated with the observed (sampled) trajectory in
phase space. Hence, the length distribution $p(l)$ of ``white''
vertical lines $l$ in the RP not containing any recurrent pair of observed
state vectors provides an empirical estimate of the distribution of RTs on the
considered orbit, which contains important information about the
\textit{dynamics} of the system under investigation.

\subsection{Recurrence network analysis}\label{sec_rna}

Recently, different approaches have been proposed for studying basic properties
of time series from a complex network
perspective~\cite{Zhang2006,Xu2008,Yang2008,Lacasa2008,Marwan2009,Donner2010IJBC}.
Many existing methods for transforming time series into network representations
have in common that they define the connectivity of a complex network -- similar
to the spatio-temporal case -- by the mutual proximity of different parts (e.g.,
individual states, state vectors, or cycles) of a single
trajectory~\cite{Donner2010,Donner2010IJBC}. Among other related approaches,
$\varepsilon$-recurrence networks (RNs) and their quantitative analysis have
been found to allow identifying transitions between different types of dynamics
in a very precise
way~\cite{Marwan2009,Donner2010IJBC,Donner2010Nolta,Zou2010,Donner2011EPJB}.

In order to construct the RN, we re-interpret the recurrence matrix $R_{i,j}$,
the main diagonal of which is removed for convenience, as the adjacency matrix
$A_{i,j}$ of an undirected complex network associated with the recorded
trajectory, i.e.,
\begin{equation}
A_{i,j}=R_{i,j}(\varepsilon)-\delta_{i,j},
\end{equation}
\noindent
where $\delta_{i,j}$ is the Kronecker delta. The vertices of this network are
given by the individual sampled state vectors on the trajectory, whereas the
connectivity is established according to their mutual closeness in phase space.
This definition of a complex network provides a generic way for analyzing phase
space properties of chaotic attractors in terms of network
topology~\cite{Donner2010,Donner2011EPJB}. However, since the network topology
is invariant under permutations of vertices, the statistical properties of RNs
do not capture the dynamics on the attractor, but its \textit{geometric}
structure based on an appropriate sampling. In this respect, we emphasize that
since a single finite-time trajectory does not necessarily represent the typical
long-term behavior of the underlying system, the resulting network properties
depend -- among others -- on the length $N$ of the considered time series (i.e.,
the network size), the probability distribution of the data,
embedding~\cite{Donner2010b}, sampling~\cite{Facchini_pre_2007,Donner2010IJBC},
etc. We choose the threshold $\varepsilon$ in such a way that the
resulting RN has a fixed edge density (recurrence rate) of $RR=0.03$ unless
otherwise stated explicitly. 

Although they primarily describe geometric aspects, the topological features of
RNs are closely related to invariant properties of the underlying dynamical
system~\cite{Marwan2009,Donner2010,Gao2009,Donner2011EPJB}. In model systems
(e.g., R\"ossler and Lorenz systems), both local and global network properties
have already been studied in great
detail~\cite{Donner2010,Donner2010b,Donner2010IJBC,Donner2011EPJB}. Among
others, two particularly interesting local measures are \begin{enumerate}[(i)]
\item the \textit{local clustering coefficient} $\mathcal{C}_v$, which
quantifies the relative amount of closed triangles centered at a given vertex
$v$ (i.e., at the associated point $\mathbf{x}_v$ in phase space) and gives
important information about the geometric structure of the attractor within the
$\varepsilon$-neighborhood of $v$ in phase space~\cite{Donner2011EPJB}, and
\item \textit{betweenness centrality} $b_v$, which quantifies the fraction of
all shortest paths in a network that include a given vertex $v$
\cite{Freeman1979}. In a RN, vertices with high $b_v$ correspond to regions with
low phase space density that are located between higher density regions. Hence,
$b_v$ yields information about the local fragmentation of the
attractor~\cite{Donner2010,Donner2010b}. Since in a complex network, the values
of $b_v$ may span several orders of magnitude, in the following we will consider
$\log b_v$ as a characteristic measure for network topology.
\end{enumerate}
\noindent In a RN, both $\mathcal{C}_v$ and $b_v$ are sensitive to the presence
of unstable periodic orbits (UPOs), but resolve complementary
aspects~\cite{Donner2010b}. Specifically, in a continuous system, it is
well-established that if a chaotic trajectory enters the neighborhood of an UPO,
it stays within this neighborhood for a certain time~\cite{Lathrop_pra_1989}. As
a consequence, states accumulate along this UPO instead of homogeneously filling
the phase space in the corresponding neighborhood (in particular if we consider
UPOs of lower period), which results in a locally reduced effective dimension
that can be quantitatively characterized by $\mathcal{C}_v$ and measures derived
from this quantity~\cite{Donner2011EPJB}.

In addition to the aforementioned vertex characteristics, several global network
measures have already proven to distinguish between qualitatively different
types of behavior in both discrete and continuous-time
systems~\cite{Marwan2009,Zou2010,Dongers2011a,Dongers2011b}. Extending these
previous results to different appearances of chaotic dynamics, we will consider four particular
measures~\cite{Newman2003,Boccaletti2006,daCosta2007} as potential candidates
for discriminatory statistics: \begin{enumerate}[(i)] \item the \textit{global
clustering coefficient} $\mathcal{C}$~\cite{watts_nature_1998}, which gives the
arithmetic mean of the local clustering coefficient $\mathcal{C}_v$ taken over
all vertices $v$, \item \textit{network transitivity}
$\mathcal{T}$~\cite{Barrat2000,Newman2001}, which is closely related to
$\mathcal{C}$ (but gives less weight to poorly connected
vertices~\cite{Donner2011EPJB}) and globally characterizes the linkage
relationships among triples of vertices in a complex network (i.e., the
probability of a third edge within a set of three vertices given that the two
other edges are already known to exist) \footnote{Note that $\mathcal{T}$ is
sometimes referred to as the (Barrat-Weigt) global clustering coefficient, often
also denoted as $\mathcal{C}$, e.g., in~\cite{Zou2010}. In order to avoid
confusion, in this work we prefer to discuss both measures separately.}, \item
the \textit{average path length} $\mathcal{L}$, which quantifies the average
geodesic (graph) distance between all pairs of vertices, and \item the
\textit{assortativity coefficient} $\mathcal{R}$~\cite{Newman2002}, which
characterizes the similarity of the connectivity at both ends of all edges in
the network (i.e., the correlation coefficient between the degrees of all pairs
of connected vertices).
\end{enumerate}
\noindent Network transitivity and average path length have already proven to
provide an excellent discrimination between complex periodic and chaotic orbits
in a two-parameter bifurcation scenario of the R\"ossler system~\cite{Zou2010}.
An analytical theory for computing the value of $\mathcal{T}$ from a known
invariant density $\rho(x)$ revealed a strong relationship to a certain concept
of generalized fractal dimensions~\cite{Donner2011EPJB}. In this respect, high
values of $\mathcal{T}$ indicate the presence of a lower-dimensional structure
in phase space corresponding to a more regular dynamics. In contrast, the
average path length behaves differently for discrete and continuous-time
dynamical systems~\cite{Marwan2009,Donner2010,Zou2010}: for maps, more regular
dynamics is characterized by low values of $\mathcal{L}$, whereas the opposite
applies to chaotic oscillators.

\subsection{Recurrence properties of phase-coherent and noncoherent R\"ossler systems}\label{sec:testcase}

As a simple continuous-time deterministic dynamical system that exhibits both PC
and NPC chaotic dynamics, we first study the behavior of the RP-based concepts
described in Sec.~\ref{sec_rts} and \ref{sec_rna} for the R\"ossler system
(Eq.~(\ref{eq_roessler})). In the following, we will use numerical simulations
of this system for various parameters $a$, obtained with a fourth-order
Runge-Kutta integrator with step width $h=0.01$. The resulting trajectories have
been down-sampled to $N=10,000$ data points with a sampling interval of $\Delta
t=0.2$, which avoids strong effects of trivial temporal correlations.

In order to illustrate qualitative differences in the behavior of the RP-based
indicators for PC and NPC dynamics, we consider the individual cases $a=0.165$
(PC) and $a=0.265$ (NPC), respectively. A part of the resulting trajectories
(projected onto the $(x,y)$-plane) is shown in Fig.~\ref{fig:ros_phasespace}.
One clearly recognizes that the oscillations of the system have a well-defined
center in the PC case, but no unique center for NPC chaos.

\begin{figure}
  \centering
  \includegraphics[width=\columnwidth]{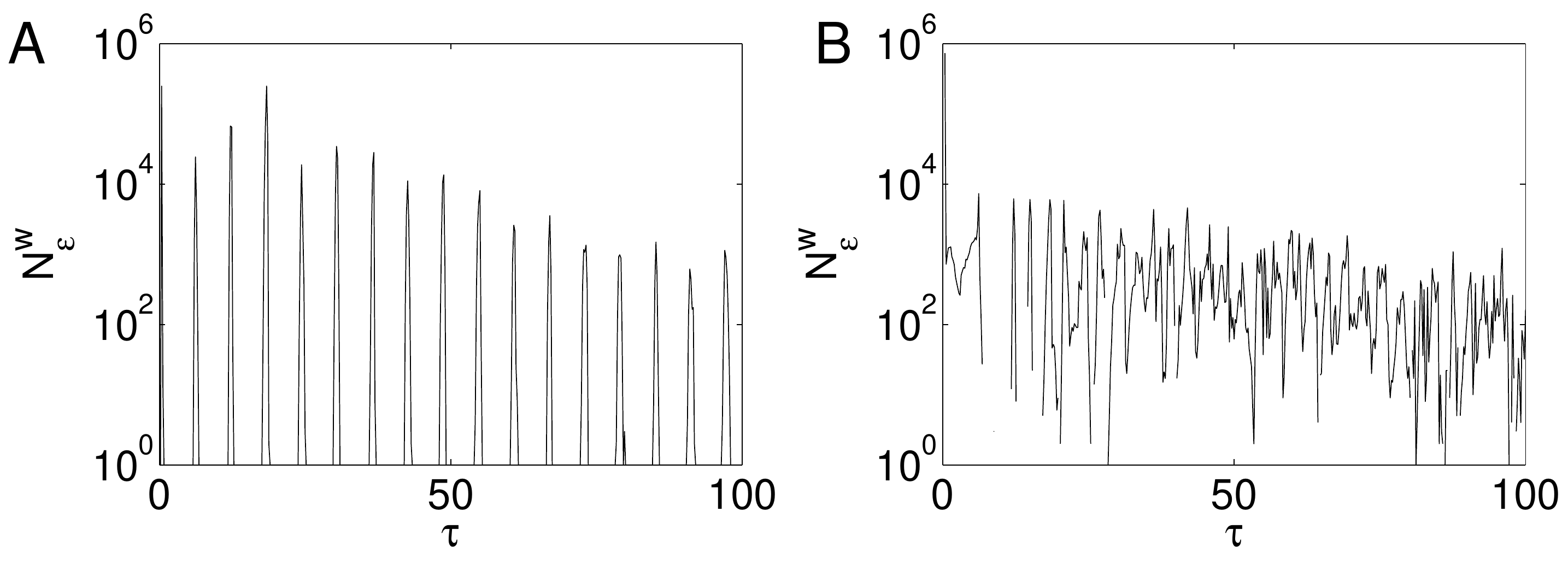}
  \caption{\small {RT distribution $p(\tau)$ with $\tau=l\Delta t$ (zoom for
  short times) for one realization of the R\"ossler system with (A) PC and (B)
  NPC chaos. The threshold $\varepsilon$ has been chosen to yield a recurrence
  rate $RR=0.03$. }
  \label{rtn_fig} }
\end{figure}

The RT distributions obtained for both examples are qualitatively different (see
Fig.~\ref{rtn_fig}). Specifically, in the PC regime the lengths of time
intervals without any recurrences are peaked around multiples of the basic
period of oscillations, with a maximum at three full periods of the
system~\cite{Thiel03} (note the logarithmic units in Fig.~\ref{rtn_fig}). This
indicates that in this regime, one distinct time-scale dominates the dynamics of
the system. In contrast, in the NPC case, the distribution becomes much more
irregular, which indicates that a multiplicity of time-scales is relevant in the
observed chaotic dynamics. However, since the complex structures in the RT
distributions have not yet been explicitly studied in previous work, it is not
\textit{a priori} clear which kind of statistical property (e.g., mean
recurrence time or the corresponding standard deviation) can be used for
distinguishing between both cases. Specifically, since the recurrence time is
related to the mean period of oscillations, its mean value varies considerably
within the different dynamical regimes as the parameter $a$ is changed.

\begin{figure}
  \centering
  \includegraphics[width=\columnwidth]{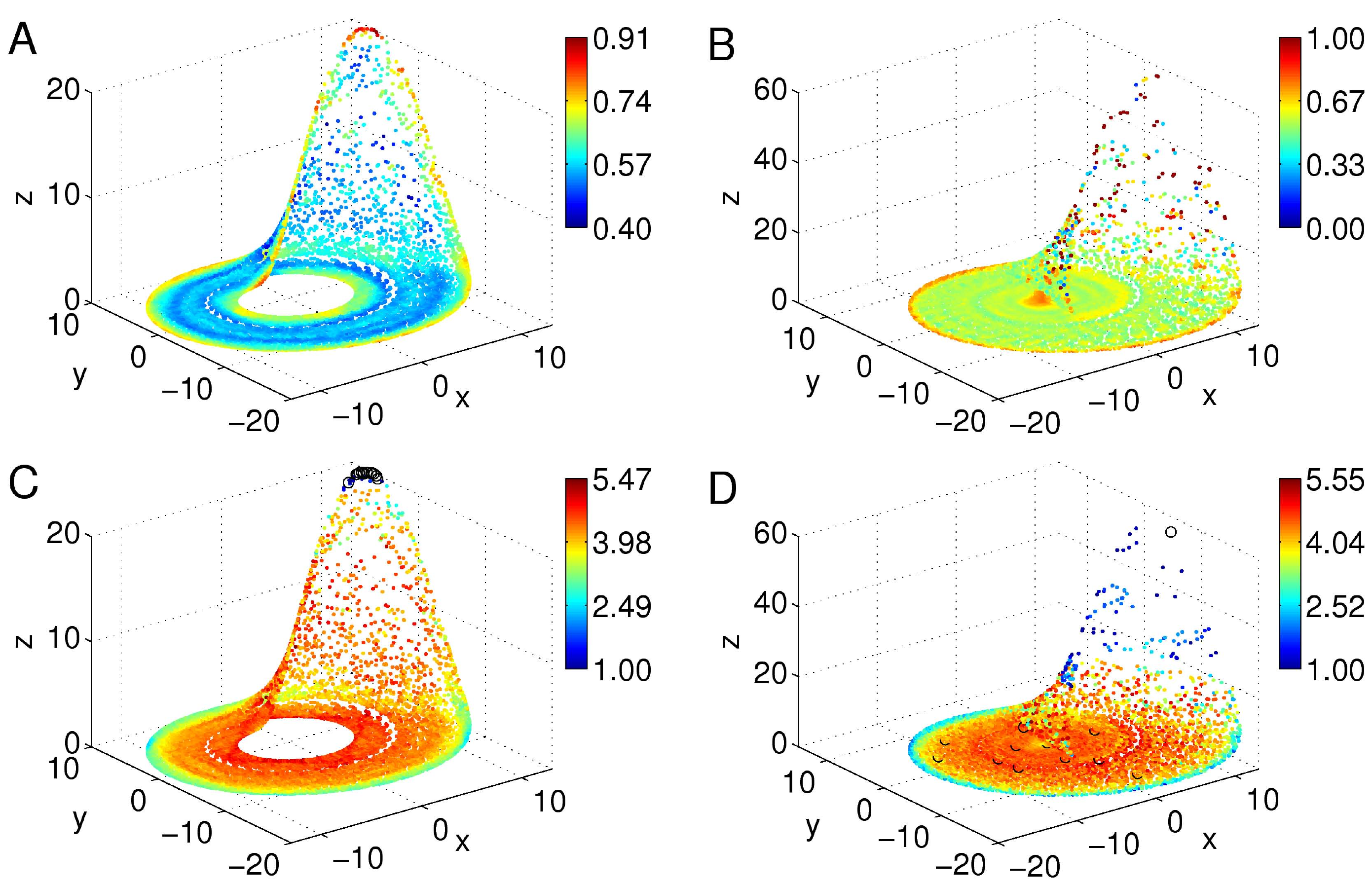}
  \caption{\small {Color-coded representations of local RN properties ((A,B) local clustering coefficient $\mathcal{C}_v$, (C,D) logarithm of betweenness centrality $\log b_v$) for the R\"ossler system with (A,C) PC and (B,D) NPC chaos ($RR=0.03$). In (C,D), black circles indicate vertices in poorly populated regions of phase space with $b_v<1$.  } \label{3d_dc_fig} }
\end{figure}

\begin{figure}[htbp]
  \centering
  \includegraphics[width=\columnwidth]{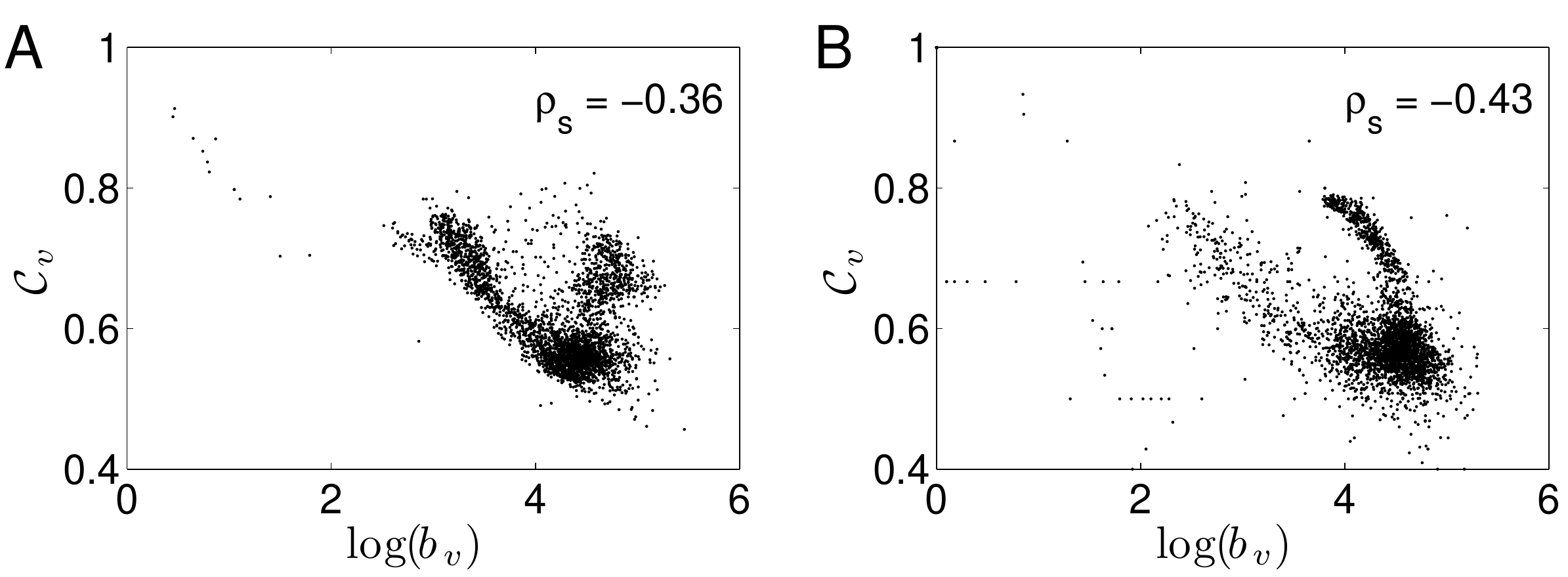}
  \caption{\small {Scatter plot between the RN measures $\mathcal{C}_v$ and $\log b_v$ for the R\"ossler system with (A) PC and (B) NPC chaos ($RR=0.03$). $\rho_s$ gives the values of the rank-order correlation coefficient (Spearman's Rho) between both quantities.} \label{dcbc_fig}}
\end{figure}

In contrast to the RT statistics, the local RN properties characterize higher-order features of the attractor geometry in phase space rather than dynamical aspects~\cite{Donner2010}. While global network properties have been recently applied for automatically discriminating between chaos and periodic dynamics in a complex two-parameter bifurcation scenario of the R\"ossler system~\cite{Zou2010}, we suggest that local properties are able to characterize even more subtle structural changes of the system. For the two considered test cases, Fig.~\ref{3d_dc_fig} shows the pattern of the local clustering coefficient $\mathcal{C}_v$ and betweenness centrality $b_v$ in phase space. It is clearly visible that both measures characterize different aspects of attractor geometry~\cite{Donner2010b}, which results in a correlation coefficient that is still significant, but not very large (Fig.~\ref{dcbc_fig}). Specifically, both measures are somewhat sensitive to the presence of UPOs, which are densely embedded in the chaotic attractor. However, while the corresponding direct relationship has been theoretically established only for $\mathcal{C}_v$ so far in terms of an effective local dimension of the attractor~\cite{Donner2011EPJB}, $b_v$ is no direct indicator for UPOs.

\begin{figure}
  \centering
  \includegraphics[width=\columnwidth]{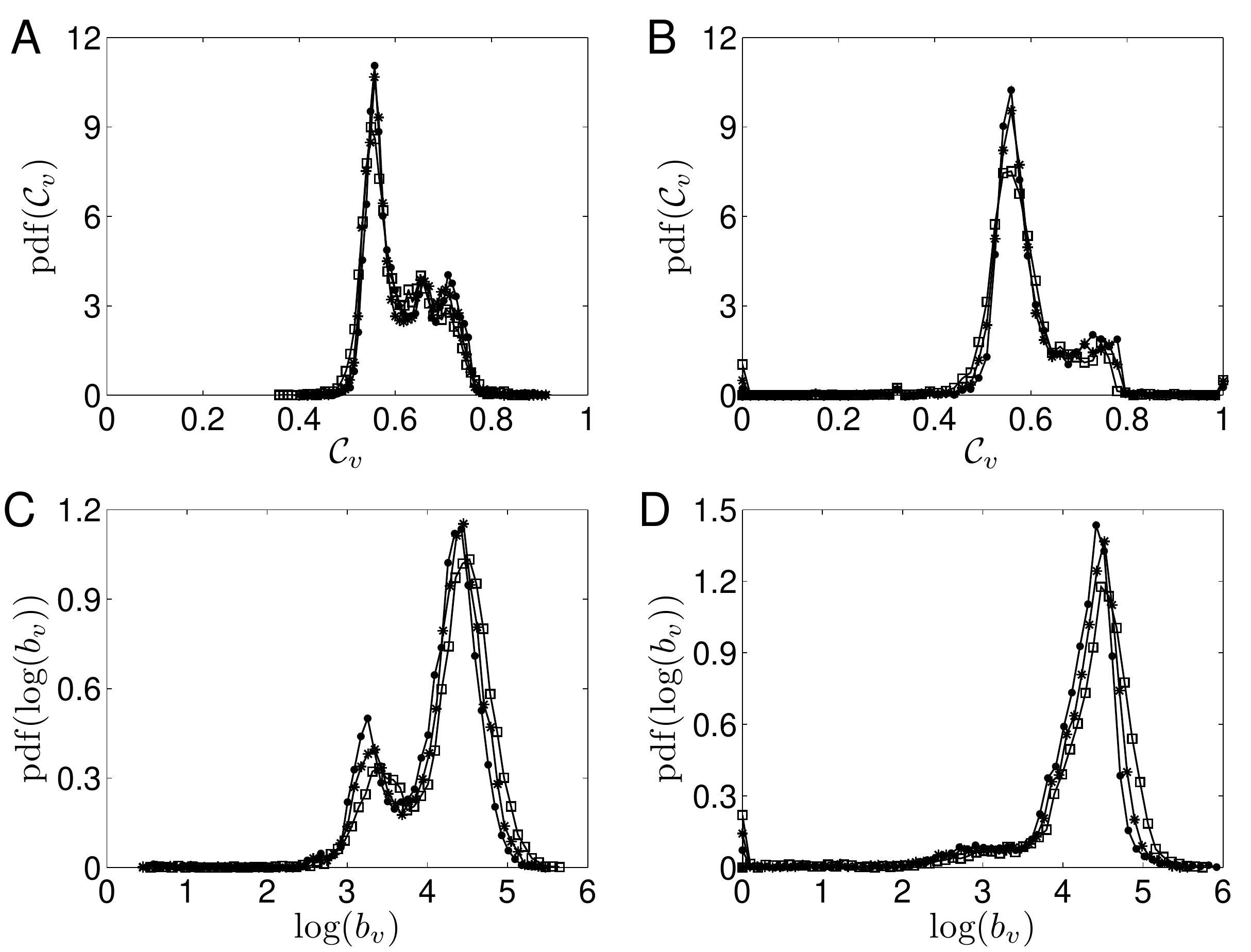}
  \caption{\small {Probability density function of the RN
  measures (A,B) $\mathcal{C}_v$ and (C,D) $\log b_v$ for the R\"ossler system
  with (A,C) PC and (B,D) NPC chaos. The different symbols represent the results
  obtained for the same trajectory with different choices of the recurrence rate
  ($RR=0.02$ ($\square$), $0.03$ ($\star$), and $0.04$ ($\bullet$)).} 
  \label{ccb_fig} }
\end{figure}

Studying the full probability distributions of both local network measures in some more detail (Fig.~\ref{ccb_fig}), we observe clear differences between PC and NPC dynamics. Specifically, all distributions are at least bimodal (which is partially related to the presence of UPOs leading to locally increased clustering coefficients), whereas the bimodality is more expressed in the phase-coherent case. Together with the general finding that the maxima of the respective distributions do not differ considerably, this result motivates considering simple statistical properties of the distributions of $\mathcal{C}_v$ and $\log b_v$ for deriving novel indices for phase coherence. We will come back to this idea in the next section.

\section{Quantifying phase coherence of chaotic oscillators}\label{sec:pcmeasures}

\subsection{Phase and frequency of chaotic oscillators}\label{phase}

In order to numerically study phase coherence of chaotic oscillators, a reasonable definition of a phase variable is usually required first. While the derivation of optimum phase variables has been recently attracted considerable interest~\cite{Kralemann2007,Kralemann2008,Schwabedal2010a,Schwabedal2010b}, we restrict our attention in this work to the standard analytical signal approach. Here, a scalar signal $x(t)$ is extended to the complex plane using the Hilbert transform
\begin{equation}
y(t)=\frac{1}{\pi}\mathcal{P.V.}\int_{-\infty}^{\infty} \frac{x(t)-\left<x\right>}{t-s}\ ds,
\end{equation}
\noindent
where $\mathcal{P.V.}$ denotes Cauchy's principal value of the integral, which yields the phase
\begin{equation}
\phi(t)=\arctan\frac{y(t)}{x(t)}.
\end{equation}

We emphasize that this definition is appropriate for oscillations with a well-defined center in the origin of the $(x,y)$-plane. Specifically, for PC dynamics, it is possible to find simple (linear) transformations of $x$ and $y$ (e.g., subtracting the mean) so that the oscillations are centered around the origin. In contrast, NPC dynamics is characterized by the non-existence of such a unique central point in phase space (cf.~Fig.~\ref{fig:ros_phasespace}B). As a result, defining the phase in the above way leads to a variable that does not monotonously increase with time. Within the framework of phase synchronization analysis, an alternative phase definition has therefore been proposed based on the local curvature properties of the analytical signal~\cite{Chen2001,Osipov2003,Kiss2005}, i.e.,
\begin{equation}
\tilde{\phi}(t)=\arctan\frac{dy(t)/dt}{dx(t)/dt}.
\end{equation}
\noindent
We note that the proper evaluation of the derivatives in the latter equation may pose substantial numerical challenges, especially in the case of (noisy) experimental data.

The instantaneous frequency of a chaotic oscillator is defined as the derivative of the phase variable with respect to time. Averaging this property over time yields the mean frequency
\begin{equation}
\omega=\frac{1}{2\pi}\left<\frac{d\phi(t)}{dt}\right>.
\end{equation}

Since in the standard Hilbert transform-based definition, the phase variable $\phi(t)$ does not necessarily increase monotonously in time, we quantify this monotonicity in order to obtain a simple heuristic order parameter for phase coherence, which we will refer to as the \textit{coherence index}
\begin{equation}
CI=\lim_{T\to\infty} \frac{1}{T} \int_0^{\infty} \Theta(-\dot\phi(t))\ dt
\label{eq:ci}
\end{equation}
\noindent
with $\dot\phi(t)=d\phi(t)/dt$.

\subsection{Traditional measures of phase coherence}\label{pdc}

The classical approach to characterizing phase coherence of chaotic oscillations is based on the second-order structure function (variogram) of the detrended phase $\Phi(t)=\phi(t)-2\pi\omega t$,
\begin{equation}
D^2_{\phi}(s)=\left<\left[\Phi(t+s)-\Phi(t)\right]^2\right>.
\end{equation}
\noindent
Averaging this property over different realizations of the same process (or, as an alternative, over different time intervals captured by the same trajectory -- note that both options can be considered equivalent as long as the system under study can be considered ergodic), one may approximately describe the dynamics of phase increments as a diffusion process~\cite{Pikovsky_Kurths_synchr,Boccaletti2004,Fujisaka2005,Zakharova2008,wickramasinghe_chaos_2010}. In this case, one obtains:
\begin{equation}
D^2_{\phi}(s)=B_1s+B_0.
\label{eq_pdc}
\end{equation}
\noindent
Comparing this with classical (stochastic) diffusion processes yields the \textit{phase diffusion coefficient} $D=B_1/2$. We note that the proper estimation of this quantity from a single trajectory may be challenging, since the detection of a proper scaling window in which the above linear relationship holds may be a nontrivial task. This is particularly true for NPC dynamics, where the appropriate definition of the phase variable $\phi$ is crucial. We note that the numerical values of $D$ depend on which of the phase definitions from Sec.~\ref{phase} is used.

As an alternative approach, in recent studies on the phenomenon of coherence resonance~\cite{Pikovsky1997,Zhou2003}, it has been suggested using the \textit{coherence factor} 
\begin{equation}
CF=\left<T\right>/\sigma_T 
\end{equation}
\noindent
(i.e., the coefficient of variation of Poincar\'e return times $T_i$, with $\left<T\right>$ and $\sigma_T$ denoting mean and standard deviation of $T$) as a measure of coherence of noise-induced oscillations. This approach can be directly transferred to the problem of distinguishing PC and NPC deterministic-chaotic oscillations~\cite{Boccaletti2004}, given a properly selected Poincar\'e section in phase space. However, we emphasize that in case of NPC chaotic oscillations, the choice of such a Poincar\'e section may be a difficult task itself.

\subsection{RP-based indicators of phase coherence}\label{rp-measures}

Since the proper estimation of the phase diffusion coefficient $D$ and coherence
factor $CF$ may be challenging, we will in the following use our results from
Sec.~\ref{sec:testcase} for defining some novel indicators for phase coherence
of chaotic oscillations based on RPs. As we have already observed, the
appearance of the RT distribution $p(\tau)$ is different for PC and NPC chaos.
Since mean RT $\left<\tau\right>=\left<\tau\right>(\varepsilon)$ and the
corresponding standard deviation $\sigma_\tau(\varepsilon)$ do not provide
sufficient results when considered separately, we suggest using the coefficient
of variation instead. This idea provides a straightforward generalization of the
coherence factor $CF$ (based on the return times with respect to a
\textit{fixed} Poincar\'e section) to a comparable measure based on the
recurrence times to arbitrary $\varepsilon$-neighborhoods of previously visited
points in phase space. Consequently, we will refer to this measure as the
\textit{generalized coherence factor}
\begin{equation}
GCF=GCF(\varepsilon)=\frac{\left<\tau\right>(\varepsilon)}{\sigma_\tau(\varepsilon)}.
\end{equation}

Complementary to this approach, we also consider measures characterizing the
properties of the associated RNs. On the one hand, we suggest that some global
network characteristics may be helpful for distinguishing between PC and NPC
chaos, as they have already proven useful for discriminating between complex
periodic and chaotic orbits~\cite{Zou2010,Donner2011EPJB}. On the other hand,
since the empirical distributions of the local RN measures $\mathcal{C}_v$ and
$\log b_v$ differ primarily with respect to their variance when comparing them
for PC and NPC chaos (Fig.~\ref{ccb_fig}), we propose using the standard
deviations $\sigma_{\mathcal{C}}$ and $\sigma_{\log b}$ as two further
alternative measures for phase coherence. In addition, it may be helpful also
considering higher-order statistics of the corresponding empirical distribution
functions, e.g., their skewness $\gamma_{\mathcal{C}}$ and $\gamma_{\log b}$.

\section{Example I: Bifurcation scenario of the R\"ossler system}\label{sec:roessler}

In order to systematically evaluate the performance of established as well as
potential new RP-based indicators for phase coherence of chaotic oscillators, we
study a part of the bifurcation scenario of the R\"ossler system
(Eq.~(\ref{eq_roessler})), where the parameter $a$ is systematically varied in
the range $[0.15,0.3]$. This parameter range comprises different kinds of
dynamics, including periodic windows and PC as well as NPC chaotic oscillations.
The transition between PC and NPC chaos occurs at $a_c\approx 0.2$, which is in
reasonable agreement with previous studies using a slightly different parameter
setting (e.g., \cite{Osipov2003}). Specifically, for $a<a_c$, the observed
chaotic attractors are always PC, whereas they are NPC for $a>a_c$. In order to
properly detect the location of periodic windows and systematically exclude them
when comparing the values of our measures for PC and NPC chaos, the largest
Lyapunov exponents $\lambda_{1,2}$ of the system are additionally
computed~\cite{Wolf}.

\subsection{Traditional and recurrence times-based measures}

\begin{figure}
  \centering
  \includegraphics[width=\columnwidth]{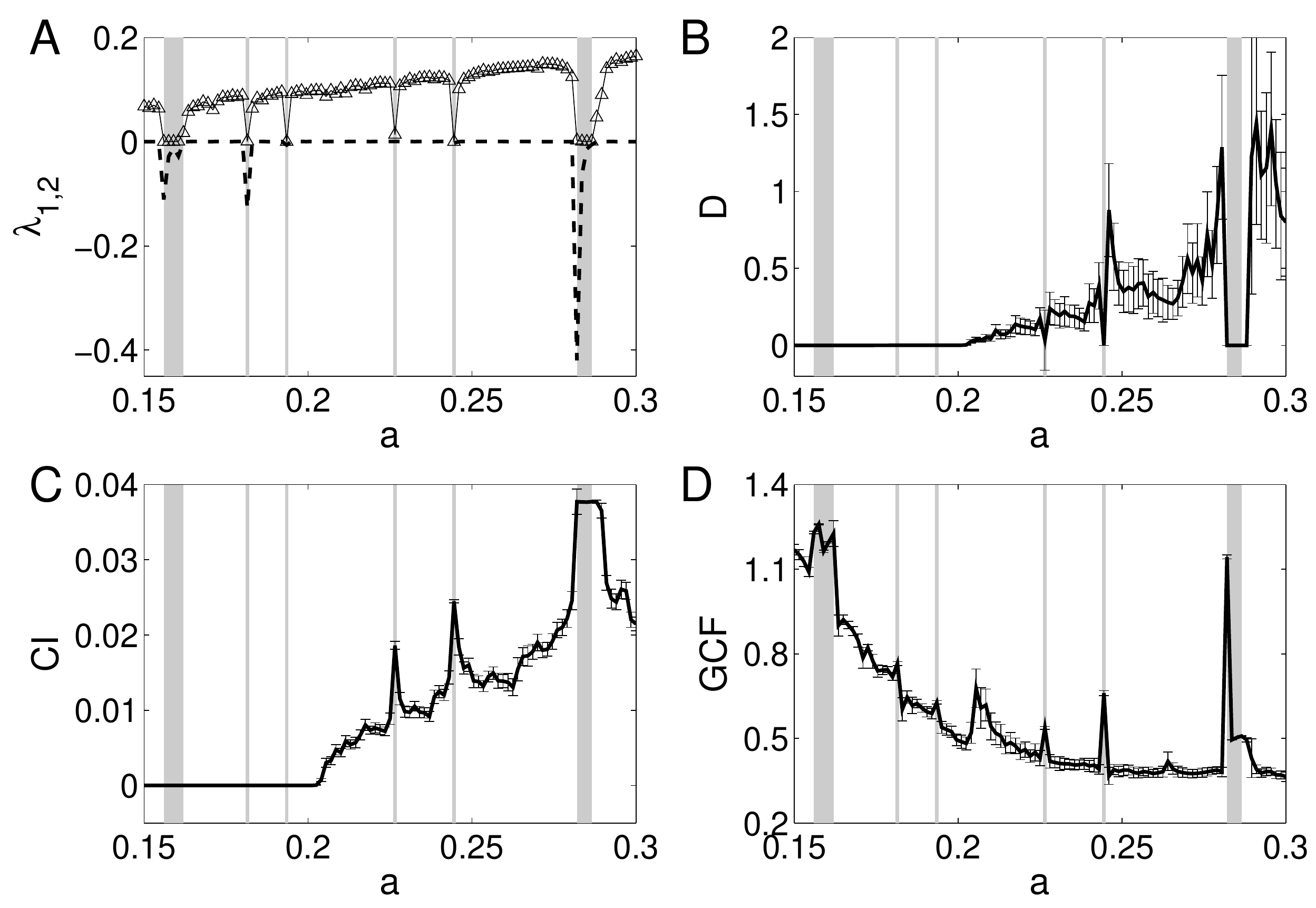} 
  \caption{\small {Behavior of different measures for phase coherence for the R\"ossler system in dependence on the parameter $a$ (error bars indicate standard deviations obtained from 100 independent realizations of the system for each value of $a$): (A) Largest Lyapunov exponents $\lambda_1$ (solid line, $\triangle$) and $\lambda_2$ (dashed line) calculated from the dynamical equations, indicating the location of periodic windows, (B) phase diffusion coefficient $D$, (C) coherence index $CI$, (D) generalized coherence factor $GCF$ ($RR=0.03$). Shaded areas indicate the presence of periodic windows evaluated by means of the largest Lyapunov exponents.} 
  \label{pc_measures} }
\end{figure}

Figure~\ref{pc_measures} displays the variation of the Lyapunov exponents
$\lambda_{1,2}$, the phase diffusion coefficient $D$, the coherence index $CI$
and the generalized coherence factor $GCF$ when the parameter $a$ is changed.
One clearly observes that the different measures are able to detect the
transition between PC and NPC oscillations at about $a=0.21$, but show different
signatures in the presence of periodic windows. Specifically, the phase
diffusion coefficient $D$ takes values close to zero ($D<10^{-3}$) in both the
periodic and PC chaotic windows, but gets much larger in the NPC chaotic regime
(Fig.~\ref{pc_measures}B). The latter observation coincides with a rather high
variance for the NPC chaotic dynamics, which is mainly due to the subjectivity
in choosing the scaling window for obtaining the linear regression parameters in
Eq.~(\ref{eq_pdc}). The coherence index $CI$ (Fig.~\ref{pc_measures}C) is zero
for $a\lesssim 0.2$, but strictly positive for higher values, including
pronounced local maxima in the periodic windows (indicating that the periodic
oscillations in these windows have no unique origin in the $(x,y)$-plane
either). In contrast, the generalized coherence factor based on the recurrence
time distributions takes very low values for NPC chaos, and higher ones for
periodic and PC chaotic windows (Fig.~\ref{pc_measures}D).

\subsection{Recurrence network measures}

\begin{figure}
  \centering
  \includegraphics[width=\columnwidth]{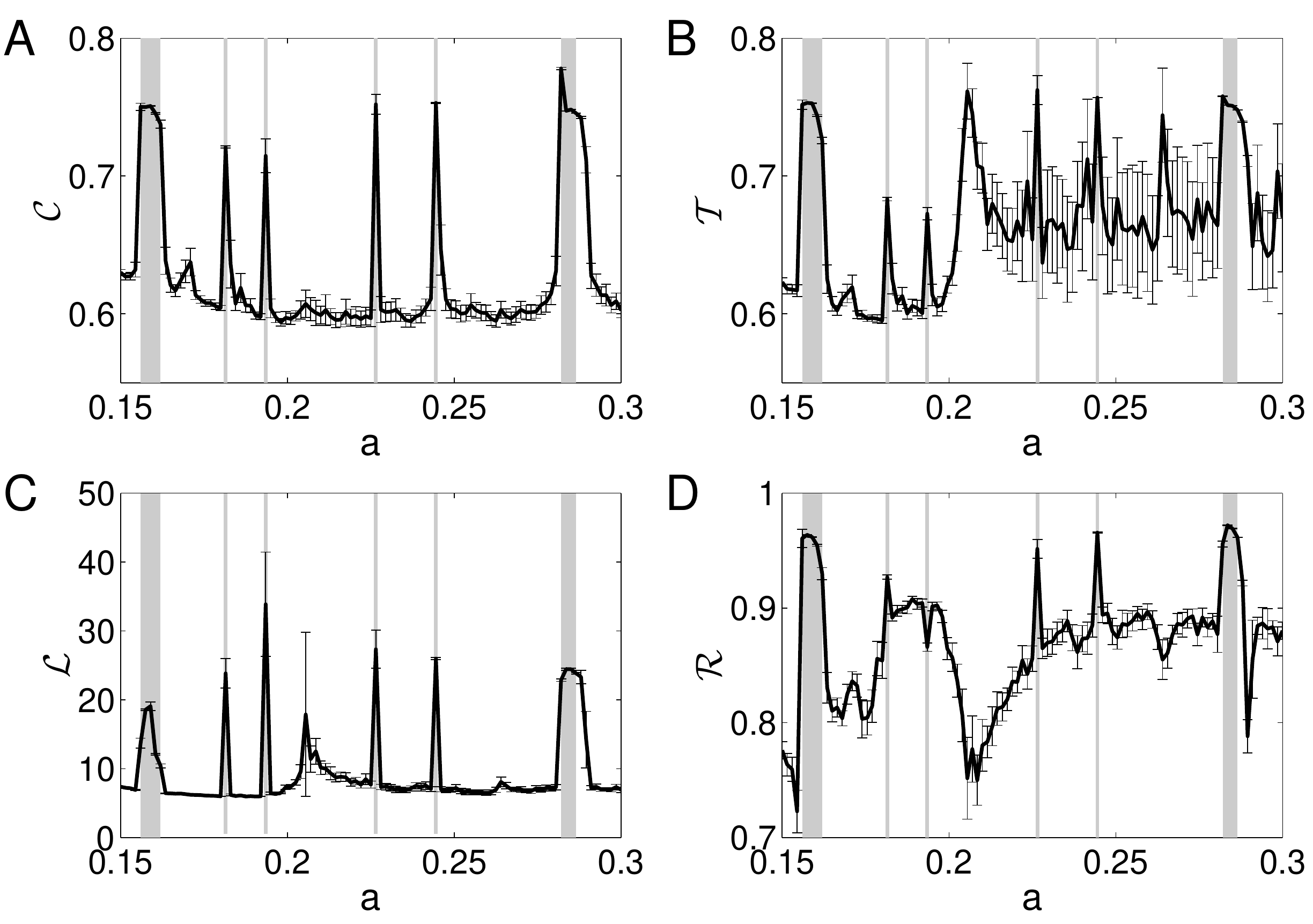} \\
  \includegraphics[width=\columnwidth]{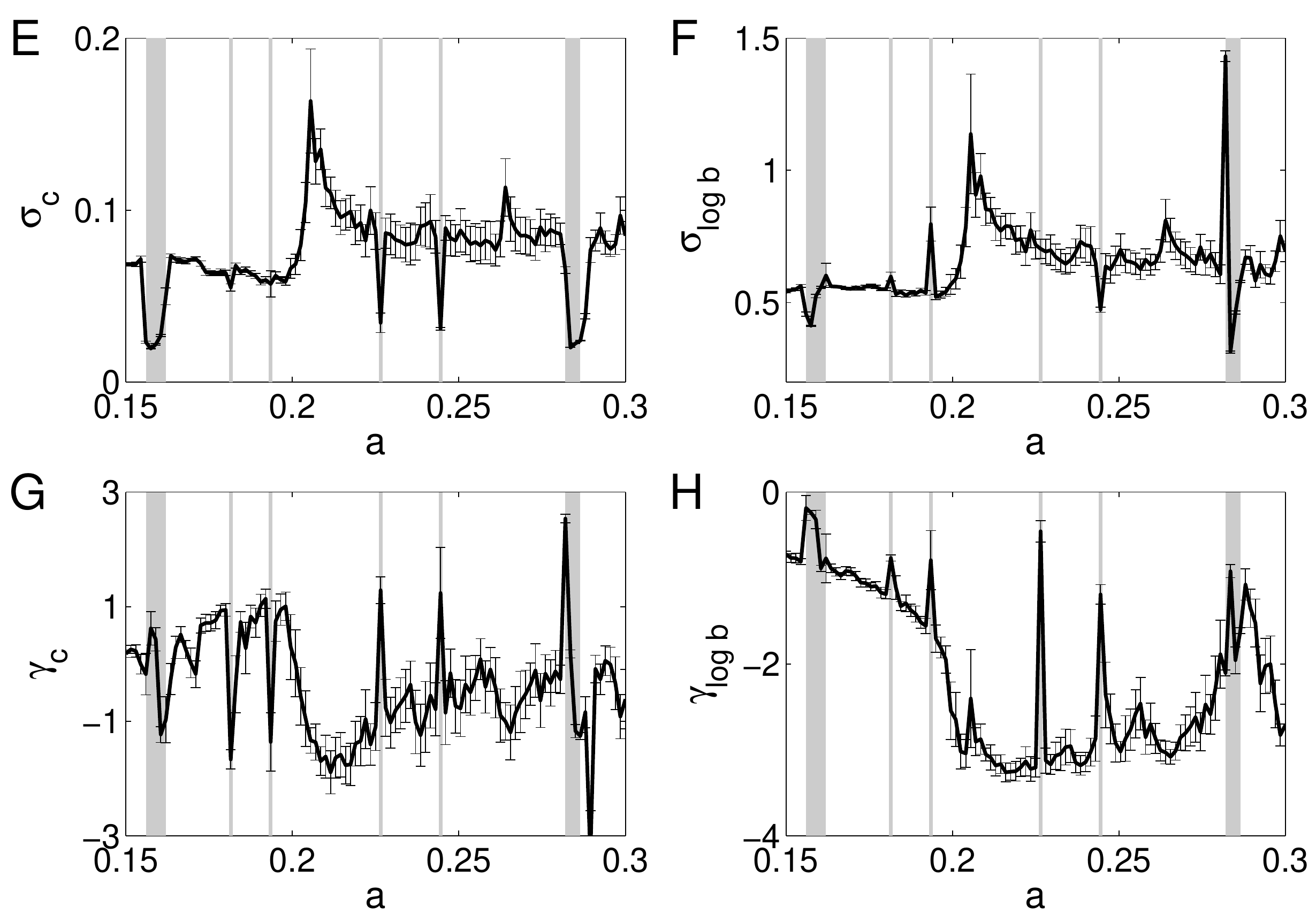}
  \caption{\small {Behavior of RN-based characteristics for the R\"ossler system in dependence on the parameter $a$ ($RR=0.03$, error bars indicate standard deviations obtained from 100 independent realizations of the system for each value of $a$): (A) global clustering coefficient $\mathcal{C}$, (B) network transitivity $\mathcal{T}$, (C) average path length $\mathcal{L}$, (D) assortativity coefficient $\mathcal{R}$, and (E,F) standard deviation and (G,H) skewness of the local clustering coefficient and logarithmic betweenness centrality ($\sigma_{\mathcal{C}}$, $\sigma_{\log b}$, $\gamma_{\mathcal{C}}$ and $\gamma_{\log b}$, respectively).} 
  \label{rn_measures} }
\end{figure}

In a similar way as described above, the values of local as well as global RN measures have been computed for realizations of the system for different values of $a$. Figure~\ref{rn_measures} shows the corresponding results. Regarding the global network characteristics, we find that the transitivity $\mathcal{T}$ has clearly higher values in the NPC regime in comparison to the phase-coherent chaos. In contrast, the assortativity coefficient $\mathcal{R}$ is clearly not capable of distinguishing both types of chaos, while a corresponding evaluation for $\mathcal{C}$ and $\mathcal{L}$ requires more detailed statistical analysis (see below). Regarding the two local RN measures $\mathcal{C}_v$ and $\log b_v$, standard deviation and skewness of both quantities show significantly higher values for NPC chaos than in the PC case, which is to be expected due to the more complex structure of the attractor in phase space. In general, the fluctuations of RN measures between different realizations obtained for the same value of $a$ are much larger in the NPC regime than for PC chaos. For the periodic windows, $\mathcal{T}$, $\mathcal{C}$ and $\mathcal{L}$ show pronounced maxima (which is consistent with previous findings~\cite{Zou2010,Donner2011EPJB}), whereas $\sigma_{\mathcal{C}}$ clearly displays local minima. In contrast, the signatures in $\mathcal{R}$ and the betweenness-based measure $\sigma_{\log b}$ are more complex. 

\subsection{Discriminatory skills of RP-based phase coherence indicators}

In order to systematically compare the discriminatory skills of all proposed RP-based measures with respect to PC and NPC chaos, we divide the set of considered values of the control parameter $a$ into three groups: one group $S_0$ representing the periodic windows (characterized by a maximum Lyapunov exponent $\lambda_1$ which does not significantly differ from zero within the numerical limits (i.e., $\lambda_1<\lambda^*=0.02$), and two groups $S_1$ and $S_2$ distinguished by whether or not the coherence index $CI$ (Eq.~(\ref{eq:ci})) does significantly differ from zero (i.e., $CI(a)< CI^*=0.001$ for PC chaos, and $CI(a)\geq CI^*$ for NPC chaos, respectively). Based on this initial discrimination, we may statistically evaluate whether or not main statistical characteristics of the distributions $p(x|S_i)$ of the different measures $x$ obtained for both groups $S_1$ and $S_2$ differ significantly. This problem can be solved by a one-way analysis of variance (ANOVA)~\cite{Montgomery2009}, with the factor being determined by two classes of values of $CI$. In order to evaluate whether the medians of some characteristic parameters in sets $S_1$ and $S_2$ differ significantly (given the variances of the empirically observed distribution functions), we perform a Mann-Whitney $U$-test~\cite{Conover1999,Hollander1999}, which can be considered as the equivalent of an $F$-test~\cite{Lomax2007} on the sets of rank numbers.

\begin{table}
\centering
\begin{tabular}{l|c|c|cl}
\hline
& PC & NPC & $P$ \\
\hline
GCF & 1.16 (0.02) & 1.17 (0.02) & 0.0177 & * \\
\hline
$\mathcal{C}$ & 0.61 (0.01) & 0.61 (0.02) & 0.0064 & ** \\
$\mathcal{T}$ & 0.61 (0.02) & 0.67 (0.03) & $2.08\times 10^{-12}$ & *** \\
$\mathcal{L}$ & 6.56 (0.78) & 8.12 (2.83) & $1.59\times 10^{-7}$ & *** \\
$\mathcal{R}$ & 0.84 (0.05) & 0.86 (0.04) & 0.2435 & --- \\
\hline
$\sigma_{\mathcal{C}}$ & 0.07 (0.01) & 0.09 (0.02) & $5.31\times 10^{-12}$ & *** \\
$\sigma_{\log b}$ & 0.56 (0.04) & 0.71 (0.09) & $1.18\times 10^{-12}$ & *** \\
$\gamma_{\mathcal{C}}$ & 0.39 (0.52) & -0.82 (0.62) & $1.33\times 10^{-11}$ & *** \\
$\gamma_{\log b}$ & -1.39 (0.65) & -2.76 (0.48) & $8.47\times 10^{-11}$ & *** \\
\hline
\end{tabular}
\caption{Mean values and standard deviations (in brackets) of the different measures for phase coherence for the considered realizations of the R\"ossler system (averages over 100 independent realizations for every value of $a$, fixed $RR=0.03$) taken over all parameter values in the PC and NPC regimes, and $P$-values of the associated $U$-test: generalized coherence factor $GCF$, global RN measures $\mathcal{C}$, $\mathcal{T}$, $\mathcal{L}$ and $\mathcal{R}$, and standard deviation $\sigma_{\cdot}$ and skewness $\gamma_{\cdot}$ of the distributions of the local RN measures $\mathcal{C}_v$ and $\log b_v$ (from top to bottom). Symbols indicate the significance of the different parameters as discriminatory statistics (---: insignificant, *: significant at 5\% level, **: significant at 1\% level, ***: significant at 0.1\% level).}
\label{tab:utest}
\end{table}

The results of our corresponding analysis are summarized in Tab.~\ref{tab:utest} and confirm our qualitative statements. Specifically, we observe that standard deviation and skewness of the distributions of $\mathcal{C}_v$ and $\log b_v$ allow a statistical discrimination of both chaotic regime with very high confidence. For the global RN measures, only network transitivity $\mathcal{T}$ performs comparably well. The average path length $\mathcal{L}$ also guarantees a reliable discrimination, whereas global clustering coefficient $\mathcal{C}$ and assortativity coefficient $\mathcal{R}$ perform clearly worse. Finally, we find that the generalized coherence factor $GCF$ based on the RT distributions in principle also allows distinguishing between PC and NPC dynamics, however, on a much lower level of significance.

\subsection{Impact of the homoclinic point on RN measures}

A detailed inspection of the previously described findings reveals two interesting aspects: First, we observe that almost all RN-based measures show an overshooting close to the transition between PC and NPC chaos (see Fig.~\ref{rn_measures}). Further investigations reveal that this effect does not result from a particular choice of sampling or the finite length of the considered realizations of the system (i.e., the presence of possibly transient behavior), but seems to be generic. Second, the behavior of the network transitivity $\mathcal{T}$ seems to contradict recent general findings on the relationship between transitivity and effective attractor dimensions~\cite{Donner2011EPJB}: the higher the effective dimension, the lower the RN transitivity. Specifically, the NPC regime has a higher dimension than PC chaos (this can be inferred from the higher maximum Lyapunov exponent indicating a higher Lyapunov dimension via the Kaplan-Yorke conjecture). Therefore, one has to expect that $\mathcal{T}$ takes higher values in the PC regime than for NPC chaos, whereas Fig.~\ref{rn_measures}B displays the opposite behavior. (In a similar way, given the known fact that for a comparable value of $\varepsilon$, periodic orbits typically have a higher average path length than chaotic ones due to the formation of geometric ``shortcuts''~\cite{Donner2010}, one would also expect $\mathcal{L}$ to be shorter in the NPC regime than for PC chaos, which is \textit{not} consistent with the results in Tab.~\ref{tab:utest}.) As we will argue below, these observations can, however, be explained in terms of the specific attractor geometry of the R\"ossler system, which is characterized by the considered RN properties.

In order to understand the aforementioned overshooting as well as counter-intuitive behavior of RN measures, recall that the chaotic attractors of the R\"ossler system are characterized by the presence of a homoclinic point at the origin. In fact, the importance of the associated homoclinic orbit for the transition between spiral-type (PC) and screw-type (NPC) chaotic oscillations has been widely recognized for the R\"ossler system as well as other chaotic oscillators with a similar transition~\cite{Gaspard1983,Gaspard1984,Kahan1999,Genesio2008,Barrio2009}. On the one hand, as the control parameter $a$ increases within the PC chaotic regime, the attractor successively grows and finally extends to the vicinity of the origin shortly before the transition to the funnel regime. On the other hand, the dynamics in the $(x,y)$-plane becomes very slow whenever a trajectory on the chaotic attractor gets close to the homoclinic point, before getting rapidly ``ejected'' out that plane following the direction of the associated unstable manifold. Thus, the growth of the chaotic attractor towards the origin has two consequences: First, the statistical properties of the distribution of ejection and re-injection ``events'' with respect to the $(x,y)$-plane changes markedly as $a$ increases towards the transition point between PC and funnel regimes, which has a distinct effect on the overall recurrence properties of the system. This is reflected by the fact that the first return maps display one distinct differentiable extremum for spiral-type chaos, but several ones for screw-type chaos~\cite{Barrio2009}. Second, due to the slow dynamics close to the homoclinic point, there is a high density of sampled points on a trajectory in the neighborhood of the origin, because the residence probability in this part of the phase space increases sharply shortly before the transition point.

\begin{figure}
  \centering
  \includegraphics[width=\columnwidth]{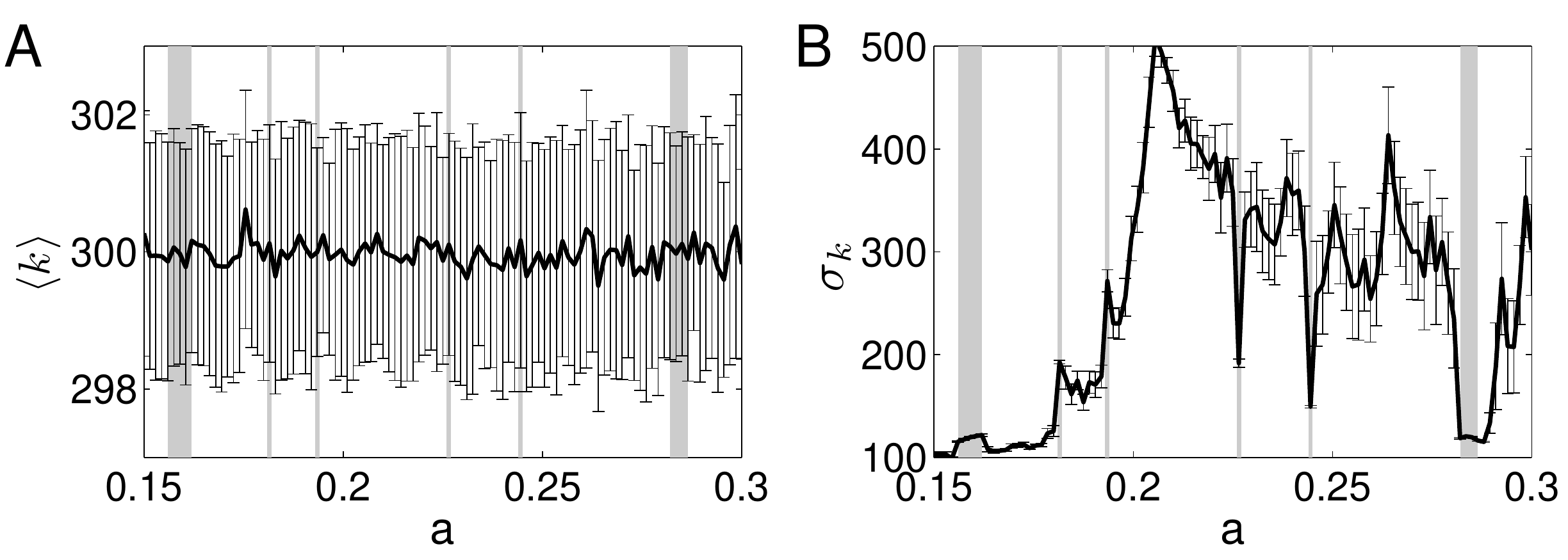}
  \caption{\small {Mean values $\left<k\right>$ (A) and standard deviations $\sigma_k$ (B) of the distribution of degree centrality $k_v$ for the RNs obtained from 100 independent realizations (error bars indicating ensemble means and standard deviations) of the R\"ossler system ($N=10,000$). The desired recurrence rate $RR=\left<k\right>/(N-1)\approx 0.03$ has been approximated by selecting the threshold $\varepsilon$ based on a Monte Carlo sampling of inter-point distances from the trajectory in order to enhance computational efficiency.} 
  \label{rn_degree} }
\end{figure}

Within the framework of RNs, the accumulation effect around the origin becomes well expressed in terms of the distribution of degree centrality $k_v=\sum_{j\neq v} A_{v,j}$, another important local network measure. Specifically, while the mean degree $\left<k\right>=(N-1) RR$ is constant when keeping the recurrence rate (edge density) fixed (Fig.~\ref{rn_degree}A), the standard deviation increases strongly shortly before the transition between both chaotic regimes (Fig.~\ref{rn_degree}B), which implies the presence of many vertices with high degree, i.e., the existence of a phase space region with a high probability density of the attractor. As a result, the local network transitivity in this distinct region increases significantly: since the neighborhoods of many vertices (in our case those located close to the origin) are densely populated (high degree), they also show a high (local) clustering coefficient ($\mathcal{C}_v\lesssim 1$, cf.~Fig.~\ref{3d_dc_fig}B). This local behavior translates into a higher global network transitivity $\mathcal{T}$ (Fig.~\ref{rn_measures}B) as well as a higher $\sigma_{\mathcal{C}}$ (Fig.~\ref{rn_measures}E). In a similar way, we can explain the overshooting of $\mathcal{T}$ and $\sigma_{\mathcal{C}}$ close to the transition point, where the variance of the degree centrality (and, hence, the density of points close to the origin) is the highest.

\begin{figure}
  \centering
  \includegraphics[width=\columnwidth]{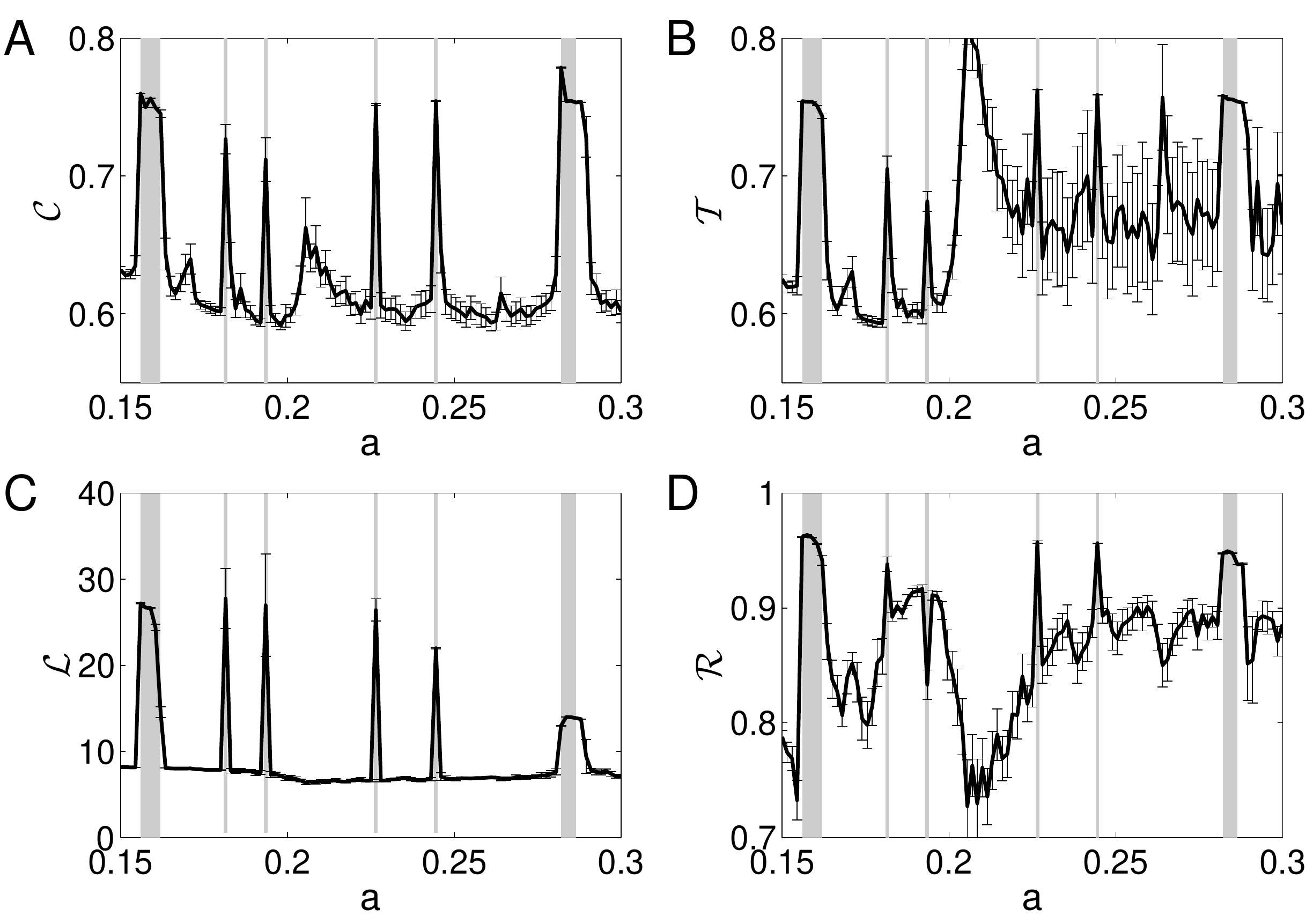}\\
  \includegraphics[width=\columnwidth]{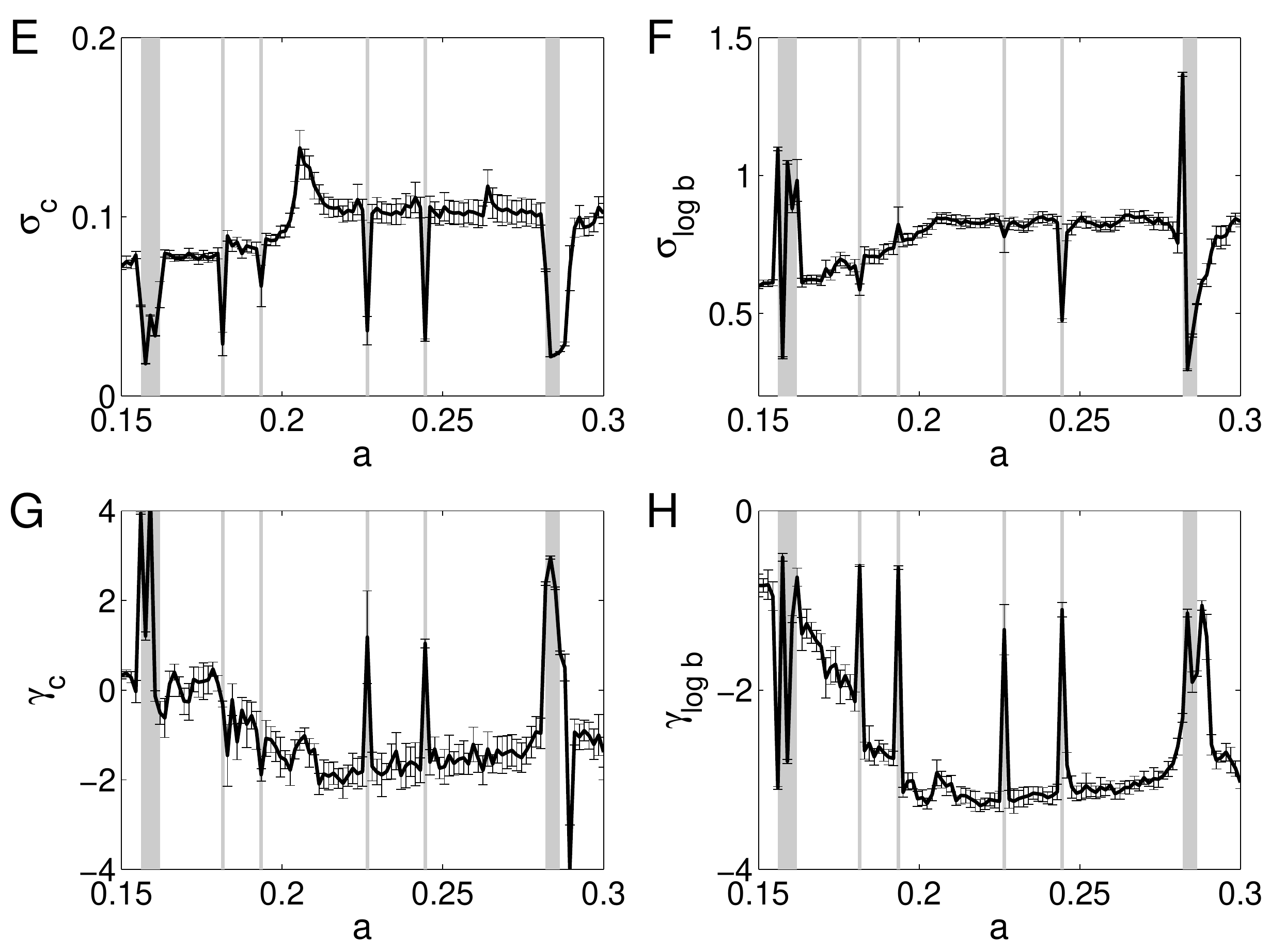}
  \caption{\small {As in Fig.~\ref{rn_measures} for a fixed recurrence threshold $\varepsilon=0.2776$ (corresponding to $RR\approx 0.03$ at $a=0.15$).} 
  \label{rn_measures2} }
\end{figure}

Regarding the effect on the path-based measures $\mathcal{L}$ and $\sigma_{\log b}$, we note that if we consider a fixed value of $\varepsilon$ (instead of a fixed $RR$) as $a$ is changed, we find no overshooting close to the transition between PC and NPC chaos (see Fig.~\ref{rn_measures2}C,F) (in a similar way, the corresponding effect is clearly reduced for $\sigma_{\log b}$ as well). Recalling the meaning of $\mathcal{L}$~\cite{Donner2010}, this observation clearly indicates that the overall size of the attractor does not change markedly close to the transition point. Moreover, $\mathcal{L}$ now takes larger values for PC chaos than in the NPC regime (Tab.~\ref{tab:utest2}), which reflects the increasing geometric complexity of the attractor. In contrast to the path-based measures, the overshooting effect on the transitivity-based measures $\mathcal{T}$ and $\sigma_{\mathcal{C}}$ persists and becomes even enhanced for the global measures $\mathcal{T}$ and $\mathcal{C}$, while it is reduced for $\sigma_{\mathcal{C}}$ and $\gamma_{\mathcal{C}}$. We emphasize that with a fixed $\varepsilon$, the recurrence rate $RR$ becomes larger when increasing $a$ close to the transition point due to the accumulation of vertices close to the origin, which could explain the aforementioned behavior. 

\begin{table}
\centering
\begin{tabular}{l|c|c|cl}
\hline
& PC & NPC & $P$ \\
\hline
$\mathcal{C}$ & 0.61 (0.02) & 0.61 (0.03) & 0.6823 & --- \\
$\mathcal{T}$ & 0.62 (0.03) & 0.68 (0.02) & $5.31\times 10^{-12}$ & *** \\
$\mathcal{L}$ & 7.70 (0.43) & 7.05 (1.01) & $9.26\times 10^{-9}$ & *** \\
$\mathcal{R}$ & 0.85 (0.05) & 0.86 (0.05) & 0.5465 & --- \\
\hline
$\sigma_{\mathcal{C}}$ & 0.08 (0.01) & 0.10 (0.01) & $3.29\times 10^{-11}$ & *** \\
$\sigma_{\log b}$ & 0.69 (0.07) & 0.82 (0.04) & $7.23\times 10^{-12}$ & *** \\
$\gamma_{\mathcal{C}}$ & -0.44 (0.71) & -1.52 (0.53) & $3.99\times 10^{-9}$ & *** \\
$\gamma_{\log b}$ & -2.15 (0.81) & -3.00 (0.38) & $1.01\times 10^{-6}$ & *** \\
\hline
\end{tabular}
\caption{As in Tab.~\ref{tab:utest} - results obtained with a fixed recurrence threshold $\varepsilon=0.2776$.}
\label{tab:utest2}
\end{table}

\section{Example II: Bifurcation scenario of the Mackey-Glass system}\label{sec:mg}

\begin{figure}
  \centering
  \includegraphics[width=\columnwidth]{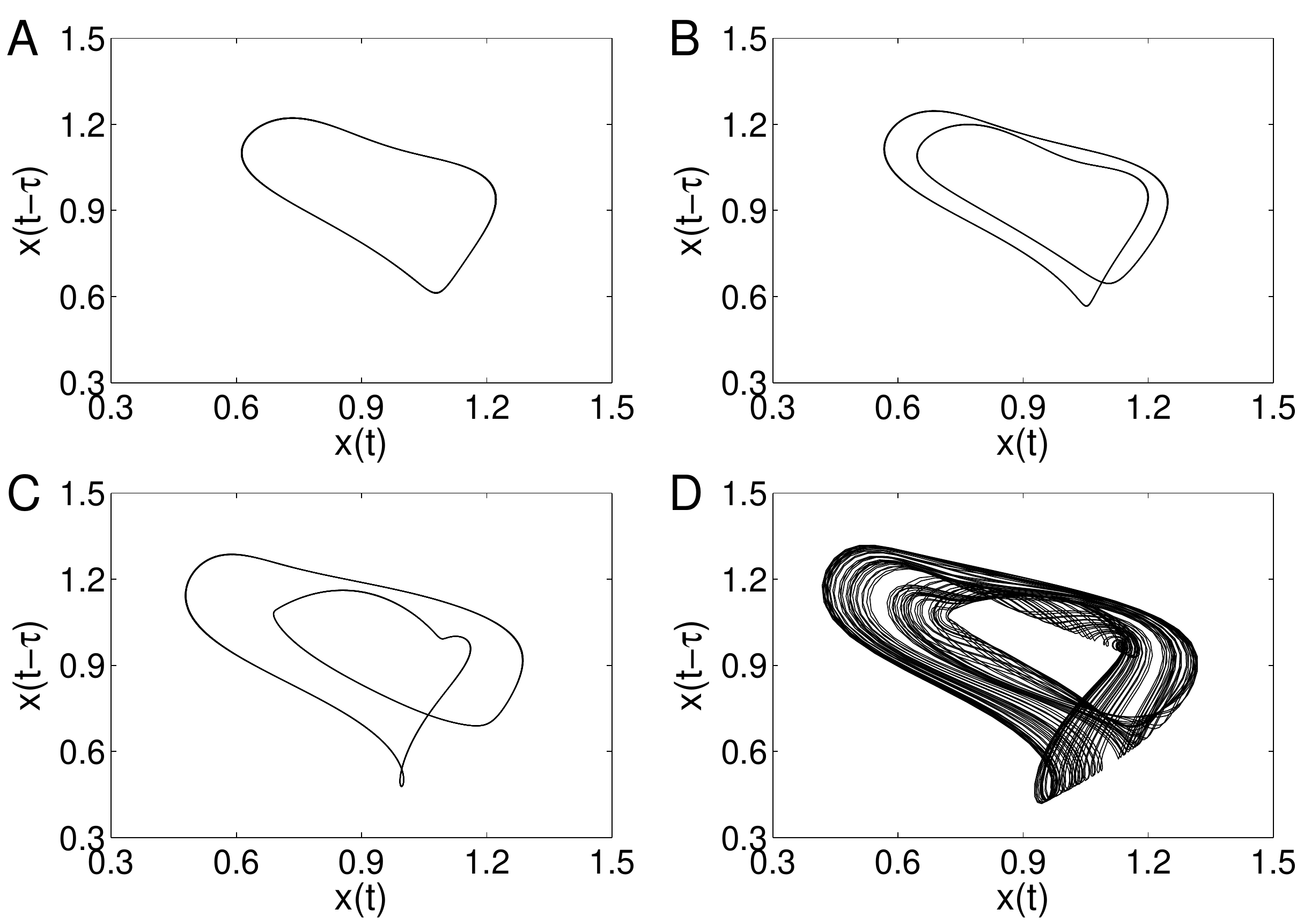} 
  \caption{\small {Phase portraits of the Mackey-Glass system (\ref{eq:mg}) for (A) $\tau=13$, (B) $\tau=13.5$ (after the period-doubling bifurcation), (C) $\tau=15.5$, and (D) $\tau=17$.} 
  \label{mg:portrait} }
\end{figure}

The scenario of a transition from spiral-type (PC) to screw-type (NPC) chaos is common to several nonlinear oscillators (for examples, see~\cite{Roessler1979,Sprott2000}). However, there are further examples for NPC chaos in other types of complex systems, especially in time-delay systems. For illustrative purposes, in the following we reexamine a part of the the bifurcation scenario of the Mackey-Glass equation~\cite{Mackey1977} 
\begin{equation}
\dot{x}(t)=\frac{0.2x(t-\tau)}{1+\left[x(t-\tau)\right]^{10}}-0.1 x(t),
\label{eq:mg}
\end{equation}
\noindent
a well-studied time-delay system, for $\tau\in[10,20]$. In this parameter range, it is known that the system undergoes several transitions between periodic and NPC chaotic solutions~\cite{Farmer1982} (see Fig.~\ref{mg:portrait}). Note that unlike the R\"ossler system, the Mackey-Glass equation describes a time-delay system, i.e., an infinite-dimensional dynamical system.

\begin{figure}
  \centering
  \includegraphics[width=\columnwidth]{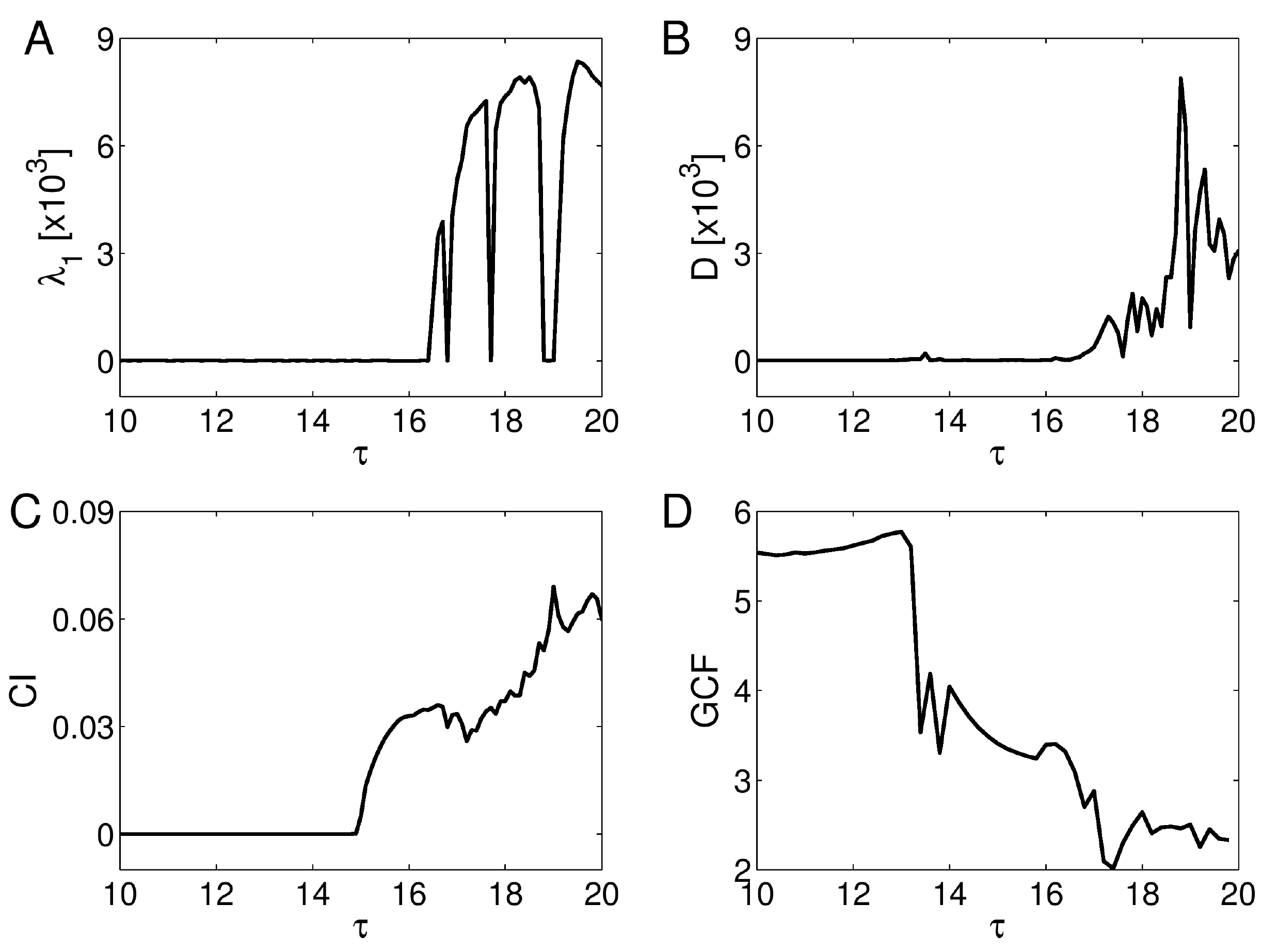} 
  \caption{\small {Behavior of different statistical characteristics for individual realizations of the Mackey-Glass system in dependence on the parameter $\tau$: (A) Largest Lyapunov exponent $\lambda_{1}$ estimated from the variational equations of a discretized version of the system with 10,000 variables representing $(x(t),x(t-\tau/9999),\dots,x(t-\tau))$, (B) phase diffusion coefficient $D$ and (C) coherence index $CI$ obtained via Hilbert transform of $x(t)$. In addition, (D) shows the mean generalized coherence factor $GCF$ obtained from 100 realizations for each value of $\tau$ ($RR=0.03$, embedding dimension 3 and delay $\tau/2$, i.e., $\mathbf{x}_i=(x(t_i),x(t_i-\tau/2),x(t_i-\tau))$).} 
  \label{fig:mg1} }
\end{figure}

Figure~\ref{fig:mg1}A shows the behavior of the maximum Lyapunov exponent with changing control parameter $\tau$. For $\tau>16$, the Mackey-Glass system switches back and forth between periodic limit-cycle oscillations ($\lambda_1=0$) and chaotic solutions ($\lambda_1>0$). However, the phase diffusion coefficient $D$ starts increasing from almost zero to non-zero (but still very small) values only at somewhat larger $\tau$ (Fig.~\ref{fig:mg1}B), pointing to a gradual loss of phase coherence with rising control parameter. We note that this finding is distinctively different from those made for the R\"ossler system, where the system undergoes a rather sharp transition from PC to NPC chaos. The behavior of the coherence index $CI$ (Fig.~\ref{fig:mg1}C) based on the standard Hilbert phase even shows a clear transition towards significantly positive values \textit{before} the establishment of the first chaotic solution. This fact is clearly related to the specific geometry of the attractor forming a small secondary loop structure after about $\tau=15$ in the $(x(t),x(t-\tau))$-plane (see Fig.~\ref{mg:portrait}C). Finally, $GCF$ (Fig.~\ref{fig:mg1}D) shows a sudden drop at $\tau>13$ (due to the presence of a period-doubling bifurcation~\cite{Glass1979} leading to marked changes in the RT distribution of the periodic solutions),  followed by a clear downward trend for further increasing $\tau$. 

\begin{figure}
  \centering
  \includegraphics[width=\columnwidth]{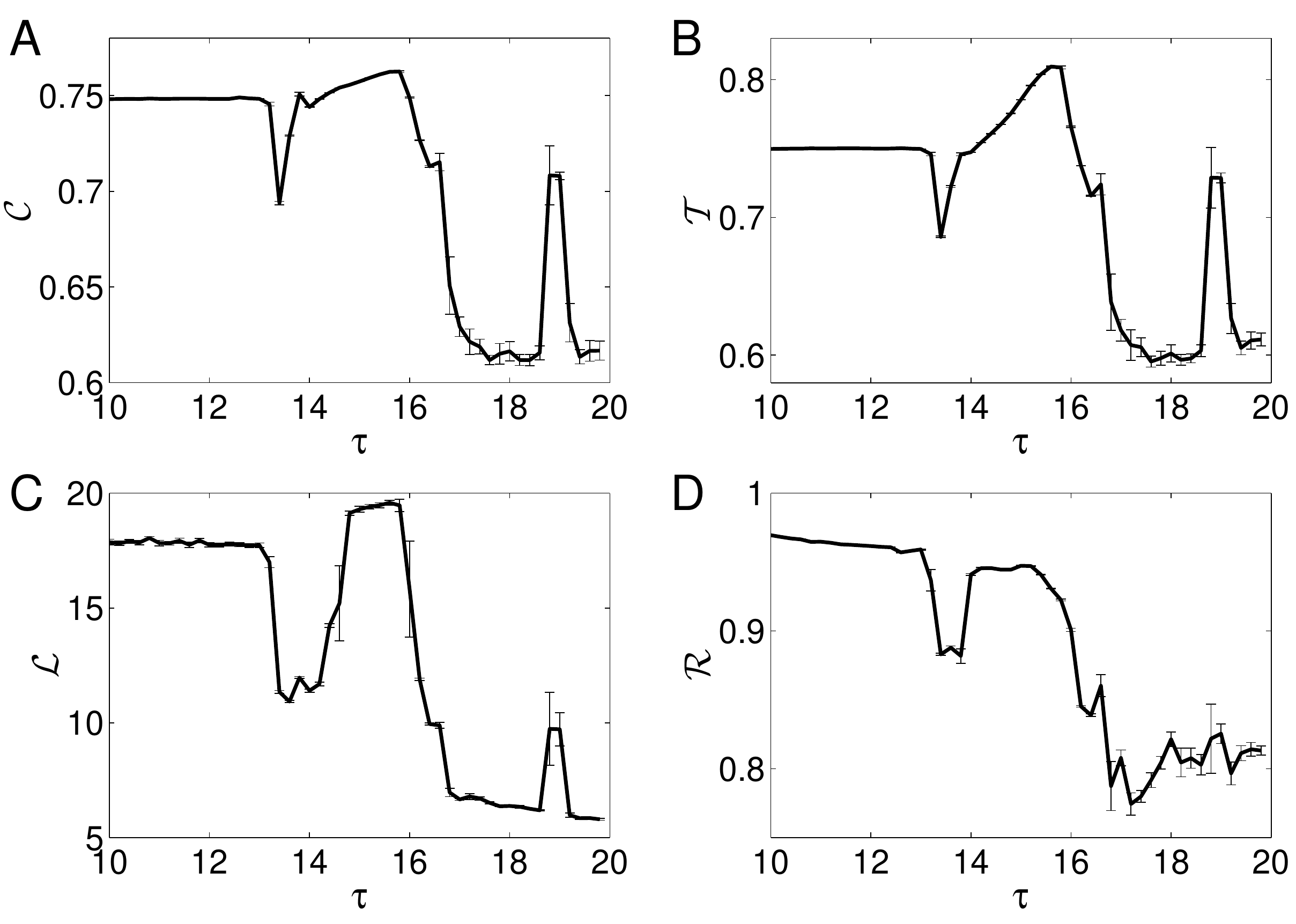} \\
  \includegraphics[width=\columnwidth]{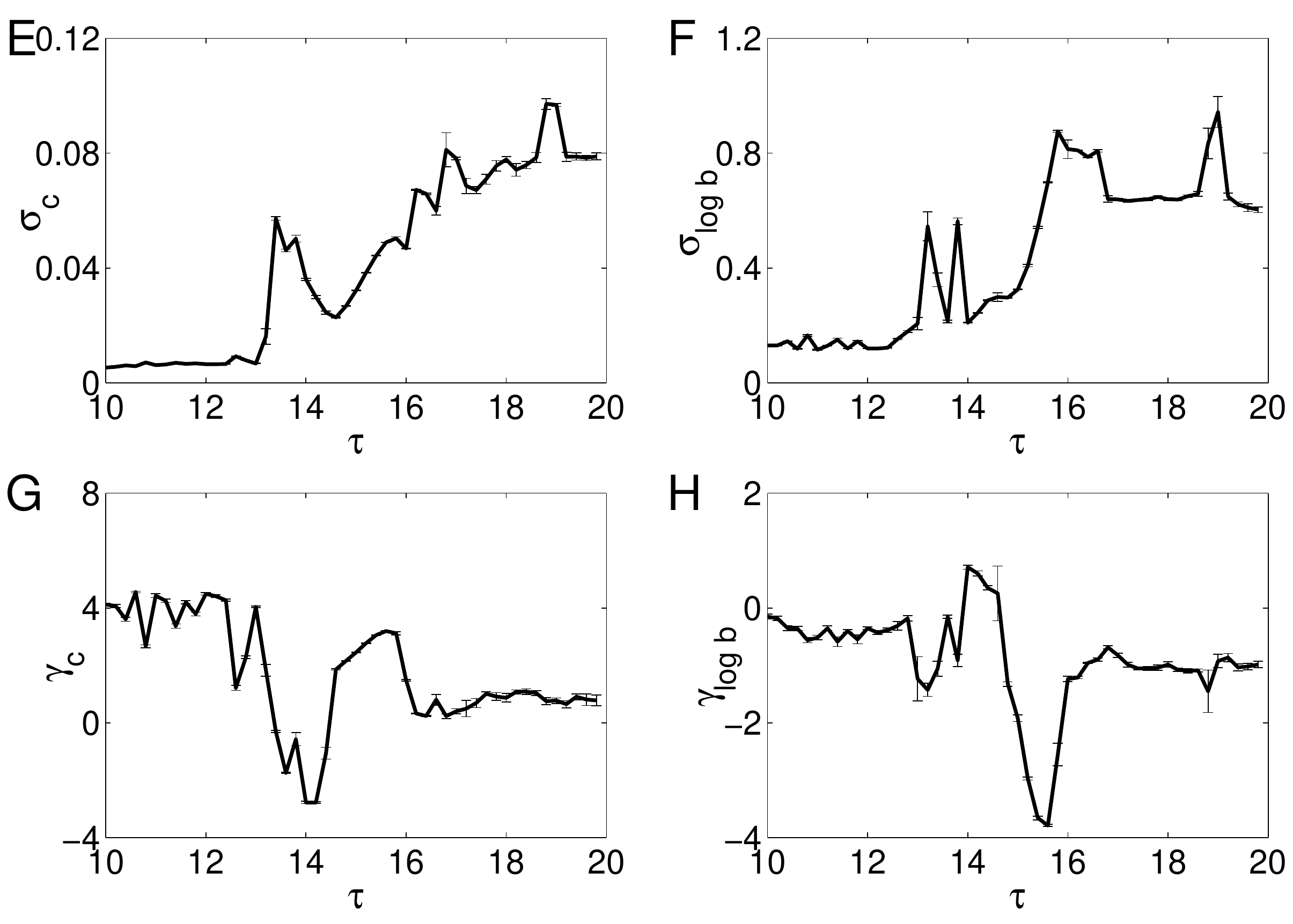}
  \caption{\small {Behavior of RN-based characteristics for the Mackey-Glass system in dependence on the parameter $\tau$ ($RR=0.03$, embedding parameters as in Fig.~\ref{fig:mg1}D, error bars indicate standard deviations obtained from 100 independent realizations of the system for each value of $a$): (A) global clustering coefficient $\mathcal{C}$, (B) network transitivity $\mathcal{T}$, (C) average path length $\mathcal{L}$, (D) assortativity coefficient $\mathcal{R}$, and (E,F) standard deviation and (G,H) skewness of the local clustering coefficient and logarithmic betweenness centrality ($\sigma_{\mathcal{C}}$, $\sigma_{\log b}$, $\gamma_{\mathcal{C}}$ and $\gamma_{\log b}$, respectively).} 
  \label{rn_mg} }
\end{figure}

The above findings are further supported by the properties of RNs resulting from
example trajectories obtained for different values of $\tau$ (Fig.~\ref{rn_mg}). As a first parameter interval of interest we consider $\tau\in[13,14]$, which is characterized by $\lambda_{1}=0$, i.e., completely periodic dynamics. Here, all network measures show a marked transition indicating structural changes of the underlying attractor corresponding to a period-doubling bifurcation. Specifically, $\mathcal{C}$ and $\mathcal{T}$ show a distinct drop from their values expected for periodic dynamics ($\mathcal{C}=\mathcal{T}=0.75$~\cite{Donner2011EPJB}), indicating the emergence of a structure of higher geometric complexity (cf. Fig.~\ref{mg:portrait}A,B). A similar marked drop is shown by $\mathcal{L}$, which is related to the emergence of geometric ``shortcuts'' after establishing the second loop of the periodic orbit. In contrast to $\mathcal{C}$ and $\mathcal{T}$, this feature persists for higher $\tau$. In addition, $\mathcal{R}$ decreases abruptly at the period-doubling bifurcation. Regarding the local network properties, we find a sharp increase in the standard deviation, and a decrease in the skewness of both clustering coefficient and log-betweenness distributions. We explain this observation by the fact that on the original single-loop limit cycle (with its rather homogeneous density), the local clustering coefficient does not vary much ($\sigma_{\mathcal{C}}\approx 0$), whereas due to the emergence of the second major loop, there exists some ``cross-over region'' within which the neighborhood of state vectors has distinctively different shape and, hence, clustering properties in the recurrence network. It is interesting to note that at the same time, the associated skewness changes its sign as the period-2 orbit successively develops ($\tau<14$) before getting back to positive values for $\tau>14$.

\begin{figure}
  \centering
  \includegraphics[width=\columnwidth]{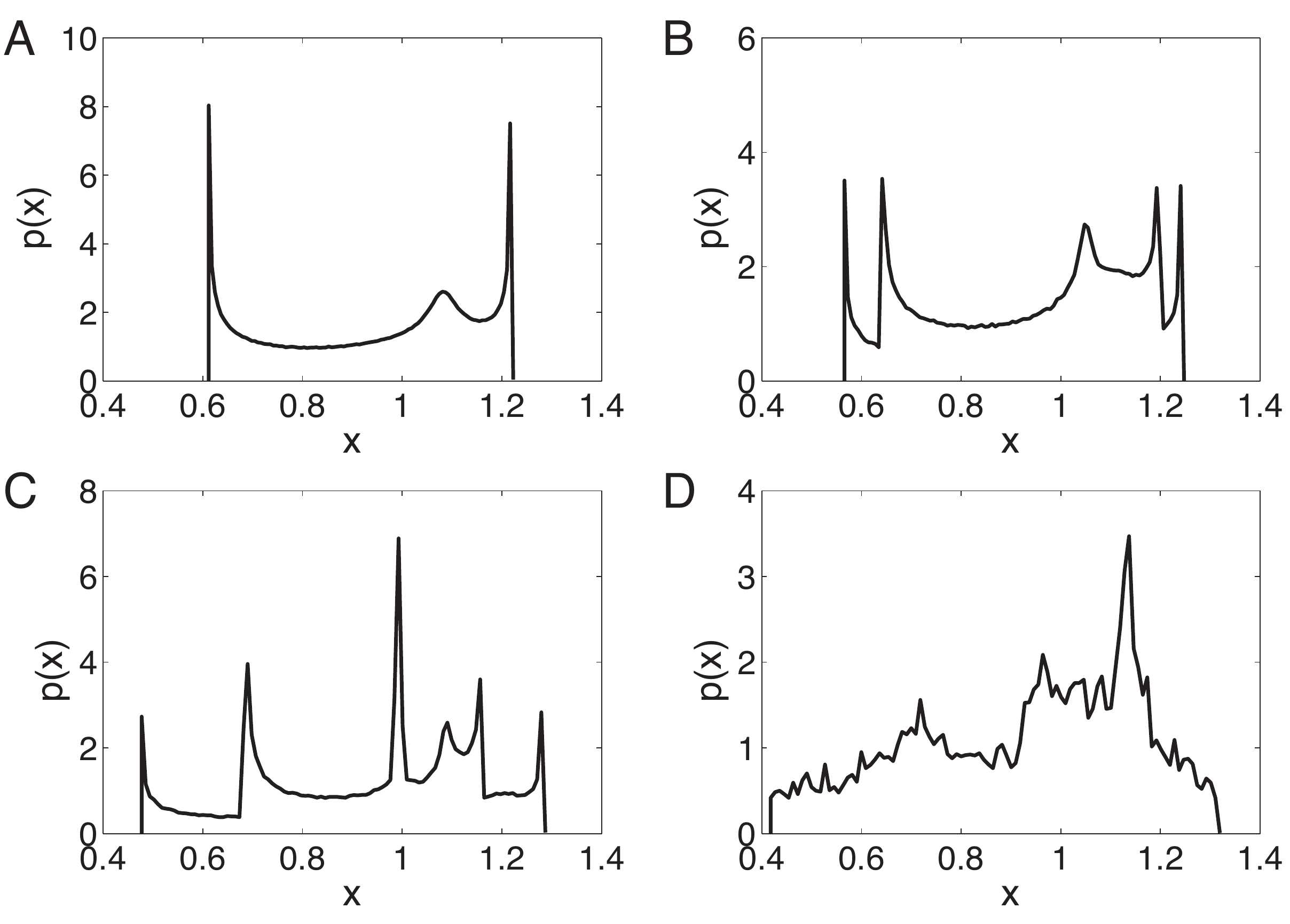} 
  \caption{\small {Estimates of the marginal density $p(x)$ of the Mackey-Glass system (\ref{eq:mg}) for (A) $\tau=13$, (B) $\tau=13.5$, (C) $\tau=15.5$, and (D) $\tau=17$. One clearly observes the period-doubling bifurcation (B) and the emergence of a cusp point and, subsequently, the secondary loop structure (C) in terms of the present local maxima.} 
  \label{mg:hist} }
\end{figure}

A second interesting parameter interval with distinct changes of the attractor
geometry is $\tau\in[14,16]$, which still refers to the periodic regime of the Mackey-Glass system ($\lambda_{1}=0$). As shown in Fig.~\ref{fig:mg1}C, at $\tau\approx 15$, the two-loop periodic orbit starts forming a cusp in the $(x(t),x(t-\tau))$-plane and, subsequently, an additional minor loop structure (Fig.~\ref{mg:portrait}C), so that the associated Hilbert phase variable does not monotonously increase anymore. In parallel to this, for $\tau>14$ both $\mathcal{C}$ and $\mathcal{T}$ increase beyond the expected values for a periodic orbit, although the maximum Lyapunov exponent clearly displays the presence of a limit cycle. This behavior results from an accumulation of probability on the trajectory close to the cusp (cf.~Fig.~\ref{mg:hist}C). Since such an accumulation has a similar effect to a recurrence network as a fixed point ($\mathcal{C}_v=1$), the overall values of the transitivity-based measures increase. The same applies to $\mathcal{L}$, where the presence of an accumulation region leads to an overall reduction of the effective $\varepsilon$ value to maintain the same recurrence rate $RR$. Similar considerations explain the increase of standard deviation and absolute value of skewness associated with the log-betweenness distribution.

While so far only bifurcations between different periodic regimes have been discussed, for larger values of $\tau$ we observe that the transition from periodic to chaotic behavior is characterized by a sharp decrease in all four considered global recurrence network measures (Fig.~\ref{rn_mg}A-D), which is consistent with the results obtained for the R\"ossler system~\cite{Zou2010} (see Fig.~\ref{rn_measures}) as well as other continuous-time dynamical systems. We note that unlike in other systems, both the periodic orbit prior to the transition point and the emerging chaotic solution are not phase-coherent (Fig.~\ref{mg:portrait}). In this respect, the Mackey-Glass system does not allow studying geometric and dynamic differences between PC and NPC chaotic solutions, but serves as an illustrative example for the presence of noncoherent (periodic and chaotic) oscillations and their impact on recurrence based characteristics.

\section{Conclusions}
In summary, we have proposed a statistical framework for characterizing
phase-coherent and noncoherent chaotic oscillations, which takes specific
geometric information about the underlying attractor into account. For this
purpose, we have utilized the recently developed concept of recurrence network
(RN) analysis. Our results demonstrate that statistical measures based on the
recurrence properties of dynamical systems do not only distinguish between
periodic dynamics and chaos~\cite{Marwan2009,Zou2010}, or quasiperiodic dynamics
and chaos~\cite{Zou_quasiperiod,Zou_chaos_2007,Zou_2008}, but also between
different appearances of chaotic dynamics characterized by phase-coherent and
noncoherent oscillations, respectively. In this spirit, RN analysis provides a
widely applicable tool for studying complex systems from a geometric point of
view, which supplements existing closely related techniques such as RQA and
recurrence time statistics which characterize complementary properties directly
related with the underlying dynamics. Specifically, besides studying systems
described by a finite set of ordinary differential equations, it has been
demonstrated that RN analysis is also applicable for describing changes in the
attractor geometry of time-delay systems such as the Mackey-Glass equation or
the piecewise linear time-delay system studied in~\cite{Senthilkumar2010NOLTA}.
However, it is not yet possible to unequivocally distinguish
between phase-coherent and noncoherent chaos exclusively based on individual
characteristics of RNs such as network transitivity. The identification of a
structural criterion for a corresponding discrimination will be a topic of our
future work.

For the R\"ossler system, we have studied the transition between spiral- and
screw-type chaos in some detail, which is common to several chaotic oscillators
and leads to a change from phase-coherent to noncoherent oscillations. The
corresponding effects on the attractor geometry and, as a result, RN
characteristics have been discussed in detail. As a particular result, we have
shown that the recently given interpretation of the RN transitivity
$\mathcal{T}$, as a measure for the effective attractor dimension, does not take
statistical effects due to a very heterogeneous distribution of residence
probability on the attractor into account, which has been largely overlooked in
previous research~\cite{Donner2011EPJB}.

In general, we find that at least for the R\"ossler system statistical
characteristics based on the distributions of local RN measures allow an equal
or even better discrimination between phase-coherent and noncoherent chaos than
some of the global network quantifiers. Moreover, both types of characteristics
outperform the studied statistical characteristics based on the recurrence time
distributions. We emphasize that RN measures probably behave so well because
they explicitly characterize geometric attractor properties in phase space
(i.e., no dynamic characteristics), which do strongly change at the transition
between phase-coherent and noncoherent chaos. In this respect, they are
particularly useful for obtaining a corresponding discrimination.

\paragraph*{Acknowledgements.} This work has been partially funded by the
Leibniz society (project ECONS) and the Federal Ministry for Education and
Research (BMBF) via the Potsdam Research Cluster for Georisk Analysis,
Environmental Change and Sustainability (PROGRESS). The authors thank Wei Zou
for providing a code for estimating the largest Lyapunov exponent of the
Mackey-Glass system, and Istvan Kiss for fruitful discussions. Complex network
measures have been calculated on the IBM iDataPlex Cluster at the Potsdam
Institute for Climate Impact Research using the software package \texttt{igraph}
\citep{Csardi2006}.


\end{document}